\documentclass[sigconf]{acmart}

\usepackage{graphicx}
\usepackage{listings}
\usepackage{xcolor}
\usepackage{booktabs}
\usepackage{multirow}
\usepackage{tabularray}
\usepackage{balance}
\usepackage{url}

\setcopyright{acmlicensed} 

\acmDOI{XXXXXXX.XXXXXXX} 

\acmISBN{978-1-4503-XXXX-X/2018/06}  

\acmConference[CCS '26]{ACM Conference on Computer and Communications Security}{2026}{The Hague, The Netherlands}
\acmYear{2026}
\copyrightyear{2026}

\definecolor{codegreen}{rgb}{0,0.6,0}
\definecolor{codegray}{rgb}{0.5,0.5,0.5}
\definecolor{codepurple}{rgb}{0.58,0,0.82}
\definecolor{backcolour}{rgb}{0.95,0.95,0.92}

\lstdefinestyle{mystyle}{
    backgroundcolor=\color{backcolour},
    commentstyle=\tiny\color{codegreen},
    keywordstyle=\color{magenta},
    numberstyle=\tiny\color{codegray},
    stringstyle=\color{codepurple},
    basicstyle=\tiny\ttfamily,
    breakatwhitespace=false,
    breaklines=true,
    captionpos=b,
    keepspaces=true,
    numbers=left,
    numbersep=5pt,
    showspaces=false,
    showstringspaces=false,
    showtabs=false,
    tabsize=2
}

\begin{document}

\title{BadQubits: An LLM-Based Framework for Static Pre-Execution Detection of Structurally Harmful Quantum Circuits}

\author{Justin Woodring}
\affiliation{
  \institution{Louisiana State University}
  \country{USA}
}
\email{jwoodr7@lsu.edu}

\author{Lamine Noureddine}
\affiliation{
  \institution{Louisiana State University}
  \country{USA}
}\email{lnoureddine@lsu.edu}

\author{Aisha Ali-Gombe}
\affiliation{
  \institution{Louisiana State University}
  \country{USA}
}\email{aaligombe@lsu.edu}

\begin{abstract}
As quantum computing moves toward commercial deployment, multi-tenant quantum clouds will increasingly receive pre-execution circuit submissions whose structural properties may be harmful to shared hardware. Some arise through intentional attacks, such as qubit-shuttling calibration exhaustion and immediate-measurement state leakage. Others arise inadvertently from poorly constructed circuits, yet their impact on shared infrastructure can be identical. In this paper, we present BadQubits, an LLM-based framework for static pre-execution detection of structurally harmful OpenQASM 2.0 circuits. The framework targets physical-execution-layer threats by analyzing submitted circuits prior to runtime, where dynamic inspection is constrained by measurement irreversibility and the exponential cost of classical quantum-state simulation.

We evaluate four code-understanding LLM architectures on a dataset of 1,500 circuits consisting of 1,000 benign programs from MQTBench~\cite{Quetschlich_2023} and 500 synthetic attack circuits derived from three documented physical-layer threat primitives. Our fine-tuned Qwen Coder 2.5 7B model achieves 92.67\% classification accuracy and 96.1\% harmful-circuit recall. Two of the four evaluated base models fail to generalize under constrained LoRA fine-tuning, indicating that architecture-aware model selection is a necessary design consideration rather than a minor tuning choice. To characterize what the detector has learned, we compare it against a bag-of-gates CNN under progressive confound removal and adversarial syntactic perturbation. The CNN's harmful-circuit recall drops from 100\% to 17\%, while the fine-tuned LLM decreases only from 96.1\% to 91.2\%. We attribute this gap to the sequential structure retained in token-level LLM inputs but discarded by histogram-based baselines. A correlation analysis further shows that model decisions track threat-defining features, specifically SWAP density and measurement timing, rather than generator-specific artifacts such as register naming. We release, to our knowledge, the first public dataset of physical-execution-layer quantum attack circuits together with generation tooling to support future research in quantum infrastructure security.

\end{abstract}

\ccsdesc[500]{Security and privacy~Malware and its mitigation}
\ccsdesc[500]{Computer systems organization~Quantum computing}
\ccsdesc[300]{Computing methodologies~Knowledge representation and reasoning}

\keywords{quantum computing security, quantum circuit analysis, large language models, QASM, structurally harmful circuits, static pre-execution detection}
\maketitle

\section{Introduction}

Quantum computing stands at the threshold of transforming computational paradigms across cryptography~\cite{Shor_1997}, optimization~\cite{quantumpotential}, drug discovery~\cite{Bravyi_2022}, and scientific simulation. As quantum processors scale from current Noisy Intermediate-Scale Quantum (NISQ) devices with hundreds of qubits toward fault-tolerant systems, quantum computing platforms are becoming accessible through larger multi-tenant cloud services provided by IBM Quantum, Amazon Braket, Azure Quantum, and Rigetti Computing. This democratization of quantum computing brings unprecedented computational capabilities to researchers, developers, and organizations worldwide, while simultaneously introducing critical security challenges that classical cybersecurity frameworks are fundamentally unprepared to address~\cite{qsurvey,faruketalthreats}. From a cybersecurity perceptive, unlike traditional systems, quantum systems exhibit unique properties that create novel attack surfaces: superposition enables probabilistic state combinations, entanglement creates non-local correlations between qubits, and measurement irreversibly collapses quantum states~\cite{qmalware}. These characteristics expose shared infrastructure to circuits we term structurally harmful: programs whose execution degrades calibration, leaks state across tenant boundaries, or exhausts routing resources, as demonstrated by qubit-shuttling attacks~\cite{shuttleexploiting}, malicious gate insertion~\cite{trojan}, and circuit obfuscation techniques~\cite{circuitobfuscation,qsidechannel}.

Despite these concerns, progress toward automated quantum circuit security analysis remains limited. Existing work has focused primarily on theoretical attack models~\cite{qsurvey}, manual analysis of specific vulnerabilities~\cite{malaq,resetoperations}, and hardware-level defenses~\cite{shuttleexploiting}. As quantum cloud platforms increasingly serve diverse user communities, this absence of detection mechanisms creates opportunities for adversaries to submit harmful circuits that target shared resources, leak information across tenants, or degrade system availability through resource-exhaustion behaviors. By design, detection of these structurally harmful circuits at the submission boundary is constrained to structural analysis alone: a pre-execution gate cannot observe intent, only code structure, and dynamic analysis is largely foreclosed by measurement irreversibility and the $O(2^n)$ intractability of classical simulation~\cite{qmalware}. Thus, for quantum shared infrastructure, static pre-execution detection is therefore not merely preferred but primarily the viable and potentially practical enforcement mechanism. However, it must reason over gate sequences, register manipulations, and measurement patterns at abstraction levels unfamiliar to classical security tools~\cite{sidechannel,gu2007denial}. 

To address this gap, we present \textbf{BadQubits}, the first LLM-based framework for static detection of structurally harmful quantum circuits prior to execution. Our approach adapts pretrained large language models~\cite{10917287} to OpenQASM 2.0 circuit representations, enabling detection of deliberately malicious, poorly designed, or inadvertently harmful programs through their structural patterns in submitted code. As concrete threat instances, we focus on two documented physical-layer attack primitives. Qubit shuttling attacks exploit quantum routing through excessive SWAP gate sequences, leading to calibration degradation and resource stress~\cite{shuttleexploiting}. Immediate measurement attacks invoke premature qubit measurements to extract residual information from prior computations or disrupt expected execution flow~\cite{resetoperations,yalereset}. Both categories leave structural traces in OpenQASM representations that are observable prior to execution. Similar patterns may also arise from poorly written circuits produced by inexperienced developers or accidental misuse. Because such circuits can impose the same operational impact on shared hardware, they belong to the same harmful-circuit risk class, making reliable discrimination from truly benign programs non-trivial. This motivates detectors that reason over structural context rather than relying on inferred intent or simple signature matching.

Our evaluation addresses three questions. (1) Can fine-tuned LLMs classify structurally harmful quantum circuits through static code analysis? (2) How do baseline and fine-tuned performance differ across code-specialized, general-purpose, and instruction-tuned LLM architectures? (3) What is the detector's dominant learned signal, and does its evasion resistance exceed that of simpler baselines? We evaluate four representative architectures: Qwen Coder 2.5 7B (code-specialized), Mistral 7B Instruct v0.3 (instruction-tuned), Seed-Coder 8B (code-specialized), and Llama 3.1 8B (general-purpose). The evaluation set is 1,500 quantum circuits combining MQTBench~\cite{Quetschlich_2023} benign samples with synthetically generated attack circuits. The fine-tuned Qwen Coder 2.5 7B model achieves 92.67\% classification accuracy with 96.1\% bad-circuit recall. In contrast, two of the four evaluated models (Mistral, Llama) collapse to degenerate classifiers under the same fine-tuning configuration. This divergence suggests sensitivity of LoRA-based adaptation to model architecture and training dynamics under class imbalance. Out of the four tested models, code understanding models exhibit more stable adaptation than general-purpose or instruction-tuned alternatives. Given the limited number of models evaluated, we interpret this as an empirical observation of the current pipeline rather than a general architectural claim. 
In summary, this work makes the following contributions:
\begin{itemize}
\item An LLM-based detection framework for structurally harmful quantum circuits at the physical-execution layer, covering both deliberately malicious and inadvertently harmful programs under a unified structural-harm threat model.
\item A multi-architecture evaluation of baseline and fine-tuned LLMs, including a controlled comparison against a bag-of-gates CNN under progressive confound removal and an adversarial syntactic perturbation. 
\item A correlation-based characterization of the LLM's dominant learned signals, designed to test whether classification is driven by threat-model-defining features or by generator-specific artifacts.
\item The first public dataset of physical-execution-layer quantum attack circuits, distinct from logical-layer adversarial datasets such as Quantum-TrojanNet and classical host-layer corpora such as MalaQ, released together with circuit-generation tooling.
\end{itemize}
\section{Background and Related Work}\label{back}
Quantum systems operate under physical and theoretical constraints that fundamentally impact security analysis: (1) \textbf{State fragility:} Quantum coherence requires operation at near absolute zero temperatures with extreme isolation from environmental interference, making quantum systems vulnerable to decoherence-based denial-of-service attacks~\cite{Bravyi_2022,gu2007denial}. (2) \textbf{No-cloning theorem:} The impossibility of creating perfect copies of unknown quantum states complicates error correction protocols and creates vulnerabilities where attackers can exploit limited redundancy~\cite{trojan}. (3) \textbf{Platform diversity:} Contemporary quantum computing employs diverse physical implementations, including superconducting qubits, trapped ions, photonic systems, and neutral atoms, each with platform-specific attack surfaces and varying susceptibility to different malicious patterns~\cite{Bruzewicz_2019,shuttleexploiting}. (4) \textbf{Measurement irreversibility:} Observing quantum states causes wavefunction collapse, destroying superposition and making runtime monitoring fundamentally incompatible with quantum computation~\cite{qmalware}. Classical malware detection draws on both static and dynamic analysis for classification and behavioral analysis. In the quantum setting, however, detecting potentially harmful or malicious circuits cannot be mitigated through runtime inspection, as quantum measurement is destructive and classical simulation of quantum behavior scales exponentially. This shifts the security burden onto static pre-execution detection to a degree that has no parallel in the classical case~\cite{resetoperations,yalereset}. 
(5) \textbf{NISQ-era limitations:} Current Noisy Intermediate-Scale Quantum devices operate with 10-1000 qubits, 0.1-1\% gate error rates, and microsecond-scale coherence times, necessitating complex calibration, error mitigation, and transpilation systems that introduce additional attack vectors~\cite{willsch2022benchmarking,transpiler}.

In addition to the aforementioned physics-based constraints, system-level factors become particularly consequential in cloud-based quantum computing environments, where circuit submission is decoupled from hardware execution and untrusted users interact with shared processors. In such settings, multi-tenant execution introduces new security risks: structurally harmful circuits can extract information from prior computations via measurement-based side channels~\cite{saki2021qubitsensingnewattack,resetoperations}, or induce resource-exhaustion behaviors that degrade calibration and monopolize processor time~\cite{shuttleexploiting,gu2007denial}. Services such as IBM Quantum, Amazon Braket, Azure Quantum, and Rigetti exemplify this model, where remote users submit circuits to shared hardware. These cloud-specific vulnerabilities motivate the need for automated pre-execution validation systems that analyze circuits before execution.

\subsection{Quantum Attack Vectors and Exploitation Techniques}
Thus, in this paper, we focus on three representative classes of structurally harmful quantum circuits documented in prior work: Qubit Shuttling Attacks~\cite{shuttleexploiting}, Immediate Measurement Attacks~\cite{saki2021qubitsensingnewattack}, and Mixed Attacks~\cite{trojan, 10483319, 8970786}. These classes capture distinct physical-layer threat mechanisms observable at the circuit level. Detailed technical descriptions are provided in Appendix~\ref{appendix:attacks}.

\subsection{Threat Model}
To reason about detection feasibility, we define the attacker capabilities under which structurally harmful circuits may be introduced. We consider adversaries that embed harmful behavior within QASM code prior to execution. The attacker is assumed to: \textbf{(1)} have access to circuit descriptions through user-level code, compromised software dependencies, or tampered transpilation pipelines; \textbf{(2)} lack direct access to the quantum processor or runtime environment, but possess sufficient knowledge to reason about the functional effects of inserted gates or modified circuit structures; and \textbf{(3)} be capable of injecting or altering instructions to induce harmful effects at the logical or physical circuit layers. This threat model reflects realistic attack surfaces in cloud-based and open-source quantum programming environments, where circuits are frequently exchanged, transformed, and executed on shared hardware.

\subsection{Related Work}
\subsubsection{Quantum Security Research}
The quantum security community has made significant progress in identifying and characterizing quantum-specific vulnerabilities. Saki et al.~\cite{qsurvey} survey security challenges across quantum hardware, software, and communication protocols. Wu and Lidar~\cite{qmalware} introduce foundational "quantum malware" concepts, establishing theoretical models of malicious quantum programs and their impact on computation integrity. Together, these foundational works show that quantum systems face security challenges fundamentally distinct from classical computing. Attack-specific research has identified concrete exploitation techniques: Saki et al.~\cite{resetoperations,saki2021qubitsensingnewattack} reveal state-leakage vulnerabilities through insecure resets and timing side channels in multi-tenant quantum clouds. Wang et al.~\cite{malaq} present MalaQ, exploiting classical frontend components to inject malicious circuits that bypass basic validation. Saki et al.~\cite{shuttleexploiting} analyze qubit-shuttling vulnerabilities in trapped-ion systems, while Stefano et al.~\cite{transpiler} show that malicious transformations during compilation can introduce backdoors. Suresh et al.~\cite{circuitobfuscation} show obfuscation techniques disguising malicious circuits as benign, and Jerry et al.~\cite{yalereset} examine measurement-based attacks. Das et al.~\cite{trojan} introduce quantum Trojan attacks that degrade the accuracy of quantum machine learning models by 23\% via malicious gate insertion. Hardware-level security research examines physical attack vectors: Bruzewicz et al.~\cite{Bruzewicz_2019} survey trapped-ion quantum computing security, identifying platform-specific vulnerabilities. Jain et al.~\cite{quantumarchitectures} analyze architectural security implications across diverse quantum computing platforms. Side-channel research~\cite{qsidechannel} demonstrates information leakage through physical observables including power consumption, electromagnetic emissions, and timing variations during quantum operations. Existing quantum security research has focused primarily on theoretical attack characterization and manual vulnerability analysis, with limited progress toward automated systems capable of analyzing arbitrary quantum circuits prior to execution. In contrast, we propose \textbf{BadQubits}, an LLM-based classification engine for detecting structurally harmful quantum circuits from their submitted OpenQASM representations.

\subsubsection{Machine Learning for Quantum Computing}
Adjacent research \cite{kremer2025practicalefficientquantumcircuit, 9892305, apak2024ketgptdatasetaugmentation} applies machine learning to quantum computing challenges such as circuit optimization, error mitigation, and algorithm discovery, showing that classical ML models can learn useful patterns from quantum circuit representations. However, this research targets performance optimization rather than security. Quantum benchmarking resources~\cite{willsch2022benchmarking,Quetschlich_2023} provide large-scale quantum circuit datasets (including MQTBench) with diverse algorithm implementations; these enable our domain-specific tokenizer training and provide benign circuit examples, but contain no attack samples or security labels. Existing quantum-security datasets target adjacent but distinct threat surfaces: Quantum-TrojanNet provides logical-layer adversarial samples targeting the integrity of quantum machine-learning models through gate-insertion attacks, and MalaQ~\cite{malaq} targets the classical host layer that manages quantum systems. No public dataset previously existed at the physical-execution layer, where attacks impact hardware calibration and cross-tenant state rather than model outputs or host-layer integrity. This paper releases the first such dataset, comprising 1500 synthetically generated physical-layer attack circuits and the generation tooling, alongside the detection framework. We regard this as a complementary resource to Quantum-TrojanNet\cite{trojan} and MalaQ\cite{malaq}, not a competing one: a full quantum-security benchmark suite should cover all three layers. Quantum-TrojanNet provides logical-layer adversarial samples targeting quantum machine-learning model integrity through embedded gate insertion, and corpora derived from MalaQ~\cite{malaq} target the classical host layer that manages quantum systems. These three corpora cover three distinct threat surfaces (classical host infrastructure, logical-layer model integrity, and physical-execution-layer hardware integrity), and should be regarded as complementary rather than competing. Detection methods targeting one layer are not in general transferable to another, and we make no claim about generalization across layers. A systematic cross-layer evaluation is an important direction but beyond our scope of this work.

\subsubsection{Classical Malware Detection and AI-based Approaches}
In the last few decades, machine learning approaches using SVMs, random forests, and neural networks have shown strong effectiveness across diverse platforms for classical malware analysis~\cite{strom2018mitre, 9848192, 10.1145/3229710.3229726, CAYIR2021102133, 6682136, 9664858, 10247307, 8628743, 6927654, dilhara2021classification, 9070608, 10.1007/978-3-030-04780-1_28, 8369054, 8703863, 5597767}. More recently, with the advancement of deep neural networks, code analysis research shows that LLMs are effective for vulnerability detection, semantic code search, and program comprehension. Models such as CodeBERT, GraphCodeBERT, and CodeGen \cite{feng2020codebertpretrainedmodelprogramming,guo2021graphcodebertpretrainingcoderepresentations, nijkamp2023codegenopenlargelanguage} learn programming semantics from large-scale corpora and transfer well to downstream tasks with limited domain-specific training data. Building on code comprehension and the versatility of LLMs, recent literature has proposed newer approaches to malware analysis. Walton et al.~\cite{10917287} demonstrate that base LLMs can achieve state-of-the-art performance in Android malware classification. Other proposed algorithms in this space also include \cite{10979936, 10877264, 11050830, Zhou_Liu_Meng_Tao_Tian_Yao_Li_Han_Chen_Yang_2025, 11050827, 11037372, ZHAO2025125546, 10757249, 10903801}. Collectively, these works validate that LLM-based approaches can learn complex security tasks, including malicious patterns, and can potentially outperform traditional ML methods. These methodologies establish important precedents for LLM-based malware detection that directly inform our quantum security approach. 
\section{BadQubits System Design}\label{design}
\textbf{BadQubits} is an LLM-based classifier designed for static pre-execution analysis of quantum circuits. It detects structurally harmful patterns, including excessive qubit-shuttling operations, immediate-measurement attacks, and mixed attack strategies. Through targeted fine-tuning, the model retains general code comprehension capabilities while improving its ability to recognize quantum-specific harmful behaviors~\cite{10917287}. The overall system design consists of three key components: (1) Preprocessing and QASM Encoding, (2) Domain-Aware Tokenization, and (3) LLM-Based Classification.

\subsection{Preprocessing and QASM Encoding}
The Preprocessing and Encoding component transforms OpenQASM 2.0 circuit inputs into a consistent intermediate representation for downstream analysis. The pipeline performs: (1) metadata sanitization that removes non-executable text (e.g., comments) to prevent non-functional content from influencing the model while preserving operational structure; (2) syntax normalization that standardizes spacing and line breaks without altering instruction order or semantics; and (3) 
circuit flattening that retains the original circuit sequence by avoiding gate decomposition, optimization, or transpilation, thereby retaining higher-level structural patterns that may be altered or obscured by compiler transformations~\cite{transpiler,circuitobfuscation,malaq}. QASM inputs are validated using Qiskit’s \texttt{qasm2} parser~\cite{qiskit2024}, which checks syntactic correctness and structural integrity of OpenQASM 2.0 code; malformed submissions are rejected prior to analysis. This deterministic, rule-based pipeline produces a consistent input representation that downstream stages can process without re-parsing.
\subsection{Domain-Aware Tokenization}
The Tokenization component serves as the linguistic bridge between raw quantum circuit code and the language model’s analytical core. Its primary function is to decompose the standardized OpenQASM code generated during preprocessing—into structured lexical units, or tokens, that preserve both the syntactic hierarchy and quantum-specific semantics of the circuit. Unlike conventional source-code tokenization, which primarily handles classical programming constructs, QASM tokenization must accurately capture the quantum computational context, including qubit indices, entanglement relationships, and gate dependencies. To achieve this objective we designed a custom tokenizer trained on the MQTBench QASM dataset~\cite{Quetschlich_2023}, which is designed to understand the OpenQASM 2.0 vocabulary—comprising quantum gates, qubit and classical registers, measurement operations, and parameterized instructions. 

A quantum circuit is represented as a sequence of QASM statements $\mathcal{C} = \{s_1, s_2, \dots, s_n\}$ where each $s_i$ is a QASM statement. Hence, the Domain-specific tokenization is:
\[
\mathcal{T} : \mathcal{C} \rightarrow \mathcal{S}, \quad \mathcal{S} = \{t_1, t_2, \dots, t_m\}, \quad t_i \in \mathcal{V}
\]
where $\mathcal\{t_1, t_2, \dots, t_m\}$ denotes individual tokens with unknown tokens mapped to $t_{\text{unknown}}$. $\mathcal{V} = \{g_i, q_j, c_k, p_l, m\}$ denotes vocabulary derived from MQTBench: $g_i$ (gate symbols), $q_j$ (qubit registers), $c_k$ (classical registers), $p_l$ (parameters), $m$ (measurements). 

This circuit representation provides the structural basis for how token sequences are processed within the BadQubits pipeline. Transforming these sequences into model-ready inputs requires management of sequence length, contextual formatting, and temporal dependencies, each of which influences how the LLM interprets circuit behavior. To address these challenges, the QASM custom tokenizer is organized into two subcomponents that ensure temporal coherence and semantic fidelity.

\subsubsection{Temporal Ordering Preservation}
Our tokenization preserves the original sequential structure of QASM code, maintaining temporal ordering and operational dependencies critical for detecting time-based attacks, such as immediate measurement exploits~\cite{resetoperations,yalereset}. By retaining execution flow, the model remains sensitive to subtle anomalies in gate sequencing and instruction timing.
Each quantum gate operation, qubit target, parameter value, and measurement instruction is tokenized to preserve semantic meaning within quantum circuit execution context, facilitating detection of attack signatures that rely on precise operation sequencing~\cite{shuttleexploiting}.
We hypothesize that harmful behavior more likely occurs in earlier circuit segments, where problematic patterns can exert influence before quantum states become entangled or measured:
\[
P(\text{attack} \mid t_1, \dots, t_k) > P(\text{attack} \mid t_{m-k}, \dots, t_m)
\]
To preserve temporal semantics, we define a timestamped token sequence:
\[
\mathcal{S}_{\text{ts}} = \{(t_1, \tau_1), (t_2, \tau_2), \dots, (t_m, \tau_m)\}, \quad \text{where } \tau_i < \tau_{i+1}
\]
Each token is embedded as:
\[
\vec{h}_i = \vec{g}_i + \vec{p}_i
\]
where \( \vec{g}_i \) is the learned token embedding and \( \vec{p}_i \) encodes the token's temporal position, ensuring execution order is retained in the model's input representation.

\subsubsection{Sequence Length Management and Chunking Strategy}
The framework processes quantum circuits using a context window of 4,096 tokens during both training and inference, ensuring compatibility with transformer architectures while maintaining efficient computational performance and inference latency. This context length covers the majority of practical quantum circuit workloads encountered in contemporary NISQ-era quantum computing applications. For variable-length circuits exceeding the 4,096-token limit, BadQubits employs a chunking strategy that divides long circuits into sequential 4,096-token segments for independent analysis:
\[
|\mathcal{S}_{\text{input}}| \leq L_{\text{max}} = 4096
\]
For circuits where $|\mathcal{S}_{\text{input}}| > L_{\text{max}}$, we partition the tokenized sequence into $k$ chunks:
\[
\mathcal{C}_{\text{chunked}} = \{\mathcal{S}_1, \mathcal{S}_2, \dots, \mathcal{S}_k\} \quad \text{where} \quad |\mathcal{S}_i| = L_{\text{max}}
\]
Each chunk is analyzed independently by the LLM, producing individual classifications $\{y_1, y_2, \dots, y_k\}$. The final circuit-level classification employs a conservative ANY-bad aggregation strategy:
\[
y_{\text{circuit}} = \begin{cases}
\text{Harmful} & \text{if } \exists i : y_i = \text{Harmful} \\
\text{Benign} & \text{otherwise}
\end{cases}
\]
where \textbf{Harmful} is defined as a class of circuit ranges that are poorly constructed or intentionally malicious, whose impact can be detrimental to the shared hardware, such as degrading calibration, leaking state across tenant boundaries, or exhausting routing resources. Whereas, \textbf{Benign} defined a good class of circuit range. This ANY-bad aggregation ensures that attacks embedded anywhere within long quantum programs are detected, even when harmful patterns appear in later circuit segments beyond the initial context window. The chunking approach enables BadQubits to handle arbitrarily long quantum circuits while maintaining the computational efficiency of fixed-length transformer models. 

\subsection{LLM-Based Classification}
We use pretrained large language models as the core classification engine because OpenQASM circuits are naturally represented as structured token sequences containing ordering, dependency, and register-interaction information that shallow feature models may discard. Sequence models are therefore well suited to reasoning over gate patterns, measurement placement, and repeated two-qubit interactions without requiring manual feature engineering.

Given a tokenized circuit, a pretrained language model $\mathcal{M}$ receives the sequence within a chat-style prompt that frames the task as binary label completion. The Qwen Coder 2.5 7B prompt template is shown in Appendix~\ref{prompt}. Classification is performed through next-token generation over a two-label vocabulary $\{\texttt{Harmful}, \texttt{Benign}\}$ rather than through a separate pooled classification head. Decoding is greedy, yielding deterministic predictions for fixed inputs and model parameters. For circuits processed through the chunking strategy, each chunk is classified independently and aggregated using the ANY-bad rule described above.

\subsubsection{LLM Fine-tuning}
In addition to the out-of-the-box pretrained base models, we fine-tune each base model to the QASM classification task. Fine-tuning adapts the pretraining prior to the quantum circuit domain while preserving the model's ability to reason about legitimate quantum instructions, data flow, and control structures. Because dynamic analysis is foreclosed in the quantum setting, all model behavior is exercised statically on token sequences without runtime feedback or quantum state simulation.

Let $\mathcal{M}_{\text{pre}}$ denote a pretrained language model. We fine-tune $\mathcal{M}_{\text{pre}}$ on a labeled dataset of QASM circuits presented inside the chat template of Section~\ref{prompt}.

\begin{itemize}
    \item \textbf{Input:} Tokenized QASM sequence $\mathcal{S} = \{t_1, t_2, \dots, t_m\}$ inside the chat-template prefix $P$.
    \item \textbf{Label tokens:} The target completion is a single token $y \in \{\texttt{Harmful}, \texttt{Benign}\}$.
    \item \textbf{Objective:} Minimize the next-token cross-entropy on the label position, with loss masked over the prefix tokens:
    
    \[
    \mathcal{L} = -\frac{1}{N}\sum_{i=1}^{N} \log p_{\mathcal{M}}\!\left(y_i \,\middle|\, P \oplus \mathcal{S}_i\right)
    \]
    \item \textbf{Adaptation:} We use 4-bit quantized LoRA adaptation on the attention and MLP projections; the base model weights remain frozen. Training details are given in Section~\ref{dataset}.
\end{itemize}

At inference time, classification is performed by greedy decoding of a single label token; we do not insert a separate classification head. This is the standard causal-LM classification formulation used by the base models under evaluation, and it preserves the pretrained token distribution of $\mathcal{M}_{\text{pre}}$ rather than discarding it in favor of a randomly initialized head.

\textbf{Class-imbalance handling.} The training distribution is 66.7\% benign, 33.3\% bad. We apply a security-oriented loss weight $w_{\text{harmful}}>w_{\text{benign}}$ at the label position, which biases the fine-tuned model toward recall at the cost of precision. The specific weighting is reported in Section~\ref{dataset}. This weighting interacts with the LoRA configuration: as discussed in Section~\ref{evaluation}, not every base model adapts stably under this combination, and the fine-tuning pipeline collapses to trivial classifiers on two of four evaluated architectures. We therefore recommend that deployments validate the fine-tuned model's behavior against a trivial-classifier baseline (all-harmful / all-benign) before committing to a model, and we make this check part of the pipeline we release. Each classified circuit is labeled \textbf{Harmful} or \textbf{Benign}. A \textbf{Harmful} label indicates patterns consistent with shuttling, immediate-measurement, or mixed attacks~\cite{malaq,shuttleexploiting,resetoperations}, or with other structurally harmful properties that degrade shared quantum resources. A \textbf{Benign} label indicates patterns consistent with legitimate quantum applications~\cite{quantumpotential}. The binary output matches the downstream security policy of admit or reject, with flagged circuits routed to human review. The system's output is a binary classification as described in Section~\ref{design}, labeling each circuit as either ``benign'' or ``harmful.'' This deterministic decision interface avoids the calibration problem of confidence-score thresholds and lets downstream security policy be expressed as a simple admit/review decision, suitable for automated submission pipelines with a human-in-the-loop fallback for flagged circuits~\cite{strom2018mitre}.

\section{Dataset, Model Selection and Configuration}\label{dataset}
This section presents a description of the dataset used for testing the baseline models and for training and testing the fine-tuned models, as well as the model selection criteria.

\subsection{Dataset}
We create a dataset of 1,500 quantum circuits comprising 1,000 benign circuits (66.7\%) and 500 attack circuits (33.3\%). 

\subsubsection{Benign Dataset}
The benign circuits encompass a wide range of established quantum algorithms sourced from the MQTBench dataset~\cite{Quetschlich_2023}, including GHZ state preparation circuits, Deutsch-Jozsa algorithms, quantum neural network implementations, Variational Quantum Eigensolver (VQE) circuits, Quantum Approximate Optimization Algorithm (QAOA) implementations, quantum error correction protocols, and various other fundamental quantum computing applications that represent legitimate use cases in contemporary quantum computing systems~\cite{quantumpotential}. The benign circuits represents 65\% of the entire dataset.

\subsubsection{Harmful Circuit Dataset}
The harmful circuits in our dataset are synthetically generated using automated generation algorithms implemented in our framework due to the absence of publicly available corpora of real-world quantum attack programs. This reflects the current state of the ecosystem: large-scale quantum cloud deployments are still emerging, and empirical datasets of harmful circuits at the physical-execution layer do not yet exist. Consequently, we construct attack instances grounded in parameterized threat models reported in prior work.

Our generation methodology follows documented attack mechanisms rather than arbitrary pattern injection. 
(1) \textbf{Shuttling attacks} are generated by inserting extended sequences of SWAP operations at varying locations within circuits, interacting with existing gate structure to induce routing pressure and resource exhaustion effects~\cite{shuttleexploiting}. Prior work reports that such patterns can degrade victim program fidelity by up to $2\times$--$63\times$, which we use as an external indicator of realism. (2) \textbf{Immediate measurement attacks} introduce premature measurement operations at different points in the circuit, including cases embedded within otherwise valid computation sequences, to exploit temporal vulnerabilities in reset behavior, aligning with the exploitation model of~\cite{resetoperations,yalereset}. For this category of attack, prior works show that even with various forms of implemented reset operations, the chance of attacker recovery of a victim's results can be up to 100\% successful, which we also use as an external indicator of realism for this kind of attack \cite{yalereset}.
(3) \textbf{Mixed attacks} on the other hand combine both mechanisms through randomized placement of SWAP sequences and measurement operations within the same circuit. 

Across all classes, the generator introduces intra-class variability through randomized qubit selection, gate placement, and circuit depth while maintaining alignment with our threat model. To reduce synthetic construction artifacts, generated circuits reuse realistic structural patterns and avoid generator-specific cues such as distinctive naming or metadata; comments are removed during preprocessing. Representative examples are provided in Appendix~\ref{examples}. To ensure that generated circuits reflect realistic program structure, we embed attacks within circuits derived from MQTBench~\cite{Quetschlich_2023}, preserving gate distributions and structural characteristics of legitimate quantum algorithms. This design prevents the dataset from degenerating into isolated patterns and forces detection methods to operate within realistic structural contexts rather than on standalone signatures.

\subsubsection{Fidelity and Dataset Validation}
We further validate the dataset by comparing structural statistics between benign and harmful circuits. Shuttling attacks exhibit significantly elevated routing activity, with a median of 114 SWAP operations (P10=17, P90=1{,}062), compared to a median of 0 in benign circuits (P10=0, P90=34), where SWAP usage is limited to specific algorithmic constructs such as QFT bit-reversal. Immediate measurement attacks exhibit a median of 0 computational gates before the first measurement (P10=0, P90=7), whereas benign circuits exhibit a median of 272 (P10=33, P90=4{,}604), with no benign circuit performing measurement before non-trivial computation. These distributions confirm that the generated attacks reflect the structural deviations defined by their underlying threat mechanisms.

Importantly, the detection problem is not trivial. Several benign algorithms exhibit structural properties that overlap with attack characteristics. For example, Deutsch--Jozsa circuits combine dense oracle operations with immediate measurement, producing patterns that resemble premature-measurement attacks at the token level. This overlap results in a 71.4\% false-positive rate on this class (15 circuits out of 21 misclassified, see Table~\ref{tab:benign_performance}, demonstrating that the task requires discriminating between structurally similar benign and harmful circuits rather than identifying isolated signatures. The classifier therefore operates solely on raw instruction sequences. Empirically, the learned signals align with the defining properties of the attack mechanisms: elevated SWAP density for shuttling attacks and measurement-order inversion for immediate measurement attacks. These observations indicate that classification is driven by threat-model-consistent structural features rather than superficial dataset artifacts.

We release the framework source code, synthetic attack circuits, generator tooling, and supplemental materials in an anonymous repository~\cite{maliciousqubits2025}, to be made public upon acceptance.

\subsubsection{Training-Testing Split and Data Imbalance Handling}
We partition the dataset (70/20/10) split with 1,080 training circuits (70\%), 300 validation circuits (20\%) and 120 test circuits (10\%) via a class-stratified split. The training partition contains 719 benign and 361 harmful circuits (66.7\%/33.3\%); the validation partition contains 198 benign and 102 harmful (66\%/34\%); the testing portion consists of 83 benign and 37 harmful (69\%/31\%). The baseline (non-fine-tuned) models are evaluated exclusively on the 300-circuit validation set to measure out-of-the-box behavior. Each model is then fine-tuned on the 1,080-circuit training set and evaluated on the same 300-circuit validation set, giving a matched-set comparison between baseline and fine-tuned performance.

\textbf{On the choice of holdout over $k$-fold cross-validation.} We use a stratified holdout split rather than $k$-fold cross-validation for the LLM evaluation. The trade-off is: $k$-fold can provide tighter variance estimates, but at approximately $k\times$ training cost. For LoRA fine-tuning of 7B--8B LLMs under 4-bit quantization, this additional cost is substantial. We selected the holdout protocol for two reasons. First, the primary objective of our study is cross-architecture comparison, and the observed performance gap between the best-performing model and lower-performing alternatives is large enough that the central ranking is unlikely to depend on modest split-level variation. Second, the complementary CNN baseline in Section~\ref{sec:cnn_comparison} is evaluated using 5-fold stratified cross-validation, allowing us to allocate additional compute toward repeated evaluation of the lower-cost model while reserving LLM resources for architecture-level comparison and robustness testing. Importantly, our adversarial perturbation and explainability analyses (Sections~\ref{sec:cnn_comparison} and Appendix\ref{sec:explainability}) depend primarily on relative behavioral trends rather than tight confidence intervals from repeated resampling. Training logs for all LLM runs are released with the repository to support future replication and extended cross-validation studies.

\textbf{Training imbalance strategy.} We address the 66.7\%/33.3\% training imbalance using security-oriented loss weighting that penalizes false negatives (missed harmful circuits) more heavily than false positives. This reflects the asymmetric cost structure discussed in Section~\ref{discussion}: in a quantum-cloud setting, a false negative permits structurally harmful code onto shared hardware, where the resulting impact may propagate across subsequent tenants, whereas a false positive incurs only human review. We note that class weighting interacts with the constrained LoRA adaptation regime and may not transfer uniformly across architectures. As reported in Section~\ref{evaluation}, two of the four evaluated base models converged to trivial or near-trivial decision behavior under the overall training configuration. Accordingly, weighting strategies should be validated per architecture rather than assumed to generalize across models.

\subsection{Model Selection}
We selected four LLM architectures spanning distinct design points: \textbf{Qwen Coder 2.5 7B} (code-specialized), \textbf{Seed-Coder 8B} (code-specialized, slightly larger parameter count), \textbf{Mistral 7B Instruct v0.3} (instruction-tuned general-purpose), and \textbf{Llama 3.1 8B} (general-purpose)~\cite{10917287}. This selection is designed to test two axes under the same fine-tuning configuration: specialization (code vs.\ general) and parameter scale (7B vs.\ 8B). As described in Section~\ref{design}, the fine-tuning process adapts each base model to binary classification over QASM circuits while preserving the pretrained prior. We use a uniform prompt template (Appendix~\ref{appendix}, Listing~\ref{prompt}) so that per-architecture variation is attributable to the model rather than the prompt. The same template is used for both the baseline (non-fine-tuned) inference and the fine-tuned inference, so the baseline-vs.-fine-tuned comparison is held on identical inputs.

\subsection{Hardware Configuration and Training Infrastructure}
The hardware configuration for this evaluation used a dual-GPU system with 47.5GB of total memory, employing 4-bit quantization and Low-Rank Adaptation (LoRA) fine-tuning to optimize model deployments across all three architectures. The evaluation framework incorporated the Unsloth optimization framework with dynamic batch sizing capabilities, enabling efficient training and inference for Qwen Coder 2.5 7B, Llama 3.1 8B, and Seed-Coder 8B models. Processing rates varied across models: Qwen achieved 0.96 samples per second for fine-tuned inference, Llama 3.1 reached 1.05 samples per second, Seed-Coder operated at 0.33 samples per second, and Mistral 7B Instruct v0.3 achieved 0.65 samples per second during evaluation.

\section{Evaluation and Results}\label{evaluation}
Using the dataset and models introduced in Section~\ref{dataset}, we evaluate BadQubits on the following objectives:
\begin{itemize}
\item \textbf{EV1} Assess the effectiveness of task-specific fine-tuning for quantum harmful circuit classification by comparing fine-tuned models against their pretrained baselines across standard performance metrics.
\item \textbf{EV2} Determine the most suitable fine-tuned architecture by comparing optimization convergence across models and characterizing the stability of the top-performing model.
\item \textbf{EV3} Compare the best-performing fine-tuned LLM with CNN baselines to evaluate robustness under confound removal and adversarial syntactic perturbation.
\end{itemize}

\paragraph{\bf Evaluation Metrics.}
We report the following metrics, all computed on the 300-circuit validation set. (1) \textbf{Accuracy} gives overall classification correctness across both classes. We report it for continuity with prior work, with the caveat that 66\% accuracy equals the benign class prior and is therefore a trivial-classifier fixed point. (2) \textbf{Precision} reflects the reliability of harmful-circuit predictions. (3) \textbf{Recall} (the harmful-circuit detection rate) is the security-critical quantity under the asymmetric cost structure of Section~\ref{discussion}. (4) \textbf{F1-score} is the harmonic mean of precision and recall, recommended over accuracy on imbalanced data~\cite{rajesh2022analysis}. (5) \textbf{Confusion matrix analysis} (FPR, FNR) provides class-level error characterization. (6) \textbf{Training loss analysis} is captured at regular intervals and is used in Section~\ref{sec:loss_analysis} to separate successful fine-tuning runs from degenerate ones.

\begin{table}[t]
\scriptsize
\centering
\caption{Baseline Model Performance - Quantum Circuit Classification}
\label{tab:Baseline}
\setlength{\tabcolsep}{3pt}
\renewcommand{\arraystretch}{1.15}
\begin{tabular}{|p{2.2cm}|c|c|c|c|c|}
\hline
\textbf{Model} &
\shortstack{\textbf{Accuracy}\\\textbf{(\%)}} &
\shortstack{\textbf{Precision}\\\textbf{(\%)}} &
\shortstack{\textbf{Recall}\\\textbf{(\%)}} &
\shortstack{\textbf{F1-Score}\\\textbf{(\%)}} &
\shortstack{\textbf{Tab}} \\
\hline
Qwen Coder 2.5 7B
& 67.00 & 66.89 & 67.00 & 56.31 & 2.67 \\
\hline
Seed-Coder 8B
& 66.00 & 43.56 & 66.00 & 52.48 & 2.90 \\
\hline
Mistral 7B Instruct v0.3
& 53.33 & 49.74 & 53.33 & 51.16 & 3.13 \\
\hline
Llama 3.1 8B
& 48.33 & 57.74 & 48.33 & 48.87 & 2.44 \\
\hline
\end{tabular}
\end{table}

\begin{table}[t]
\scriptsize
\centering
\caption{Baseline Model Confusion Matrix Analysis}
\label{tab:confusion_summary_baseline}
\setlength{\tabcolsep}{3pt}
\renewcommand{\arraystretch}{1.15}
\begin{tabular}{|p{2.2cm}|c|c|c|c|c|c|}
\hline
\textbf{Model} &
\shortstack{\textbf{True}\\\textbf{Benign}} &
\shortstack{\textbf{False}\\\textbf{Harmful}} &
\shortstack{\textbf{True}\\\textbf{Harmful}} &
\shortstack{\textbf{False}\\\textbf{Benign}} &
\shortstack{\textbf{FPR}\\\textbf{}} &
\shortstack{\textbf{FNR}\\\textbf{}} \\
\hline
Qwen Coder 2.5 7B
& 195 & 3 & 6 & 96 & 1.5\% & 94.1\% \\
\hline
Seed-Coder 8B
& 198 & 0 & 0 & 102 & 0\% & 100\% \\
\hline
Mistral 7B Instruct v0.3
& 142 & 56 & 18 & 84 & 28.3\% & 82.4\% \\
\hline
Llama 3.1 8B
& 78 & 120 & 67 & 35 & 60.6\% & 34.3\% \\
\hline
\end{tabular}
\end{table}

\subsection{EV1 - Model Effectiveness}
\subsubsection{Baseline Models}\label{baseline}
As Table~\ref{tab:Baseline} shows, the baseline models achieve accuracy values ranging from 48.33\% (Llama) to 67\% (Qwen), with Mistral at 53.33\% and Seed-Coder at 66\%. All models show poor precision/recall balance and low F1-scores. The confusion matrix in Table~\ref{tab:confusion_summary_baseline} further shows that all of the models except Llama classified the majority of samples as benign. This shows that both the general-purpose models and the code-generation models, which have primarily been trained on classical programs, struggle significantly with interpreting quantum circuits. As a result, they tend to misclassify harmful circuits as benign, highlighting a gap in quantum-specific reasoning capabilities. The parameter count (7B vs. 8B) did not influence performance, suggesting that architectural type, not model size, drives the observed differences. 

\begin{table}[t]
\scriptsize
\centering
\caption{Fine-tuned Model Performance - Quantum Circuit Classification}
\label{tab:finetuned}
\setlength{\tabcolsep}{3pt}
\renewcommand{\arraystretch}{1.15}
\begin{tabular}{|p{2.2cm}|c|c|c|c|c|}
\hline
\textbf{Model} & 
\shortstack{\textbf{Accuracy}\\\textbf{(\%)}} & 
\shortstack{\textbf{Precision}\\\textbf{(\%)}} & 
\shortstack{\textbf{Recall}\\\textbf{(\%)}} & 
\shortstack{\textbf{F1-Score}\\\textbf{(\%)}} & 
\shortstack{\textbf{Processing}\\\textbf{(s/sample)}} \\
\hline
Qwen Coder 2.5 7B 
& \textbf{92.67} & \textbf{92.44} & \textbf{92.67} & \textbf{90.87} & 1.04 \\
\hline
Mistral 7B Instruct v0.3
& 66.00 & 43.56 & 66.00 & 52.48 & 1.54 \\
\hline
Seed-Coder 8B
& 48.00 & 71.93 & 48.00 & 43.90 & 3.01 \\
\hline
Llama 3.1 8B
& 34.00 & 11.56 & 34.00 & 17.25 & 0.95 \\
\hline
\end{tabular}
\end{table}

\begin{table}[t]
\scriptsize
\centering
\caption{Fine-tuned Model Confusion Matrix Analysis}
\label{tab:confusion_summary_finetuned}
\setlength{\tabcolsep}{3pt}
\renewcommand{\arraystretch}{1.15}
\begin{tabular}{|p{2.2cm}|c|c|c|c|c|c|}
\hline
\textbf{Model} & 
\shortstack{\textbf{True}\\\textbf{Benign}} & 
\shortstack{\textbf{False}\\\textbf{Harmful}} & 
\shortstack{\textbf{True}\\\textbf{Harmful}} & 
\shortstack{\textbf{False}\\\textbf{Benign}} & 
\shortstack{\textbf{FPR}\\\textbf{}} & 
\shortstack{\textbf{FNR}\\\textbf{}} \\
\hline
Qwen Coder 2.5 7B 
& \textbf{180} & \textbf{18} & \textbf{98} & \textbf{4} 
& \textbf{9.1\%} & \textbf{3.9\%} \\
\hline
Mistral 7B Instruct v0.3
& 198 & 0 & 0 & 102 
& 0\% & 100\% \\
\hline
Seed-Coder 8B
& 48 & 150 & 96 & 6 
& 75.8\% & 5.9\% \\
\hline
Llama 3.1 8B
& 0 & 198 & 102 & 0 
& 100\% & 0\% \\
\hline
\end{tabular}
\end{table}

\subsubsection{Fine-tuned Models}\label{finetuned}
The results, as shown in Table \ref{tab:finetuned}, indicate that the Qwen Coder 2.5 7B model achieved classification performance with 92.67\% accuracy, correctly classifying 272 out of 300 quantum circuits in the evaluation dataset. The model demonstrates precision (92.44\%), a balanced recall (92.67\%) and an F1-score (90.87\%), indicating performance across both benign and harmful circuit categories. On the other hand, the performance differential between models is striking: Mistral 7B Instruct v0.3 achieved 66\% accuracy, Seed-Coder 8B achieved only 48\% accuracy despite its larger parameter count, while Llama 3.1 8B showed severely low performance with 34.00\% accuracy, essentially classifying all circuits as harmful. Notably, as shown in the confusion matrix in Table \ref{tab:confusion_summary_finetuned}, the fine-tuned Mistral model exhibited a concerning pattern, classifying all test samples as benign, suggesting overfitting to the majority class. 

\textbf{Why fine-tuning collapses two of four models.} Two of the four fine-tuned models collapsed to degenerate classifiers: Mistral 7B defaulted to all-benign prediction and Llama 3.1 8B defaulted to all-harmful prediction, with Seed-Coder 8B showing a milder version of the same failure (75.8\% FPR). Because the dataset is class-imbalanced at a 2:1 benign-to-harmful ratio, always predicting benign yields 66\% accuracy by construction, so Mistral's 66\% accuracy is not a mediocre result but evidence of full classifier collapse. This collapse is likely driven by trivial-classifier fixed points in the LoRA adaptation landscape, where constrained capacity (4-bit quantization, small LoRA rank, $\sim$400 positive samples) leaves the gradient signal too weak to escape always-positive or always-negative solutions. While the training objective penalizes this behavior through weighted loss, a naive sample-counted accuracy evaluation can still surface numbers that appear acceptable, masking a fully collapsed model. Qwen's code-specialized pretraining appears to provide a strong enough initialization to avoid this.

\subsubsection{Comparative Model Performance Analysis: Fine-tuned vs. Baseline}
The results for baseline and fine-tuned models indicate a performance difference between the two. The fine-tuned Qwen model shows a 25.67 percentage point improvement over its baseline version (92.67\% vs. 67.00\%), with improvements in precision (from 66.89\% to 92.44\%) and F1-score (from 56.31\% to 90.87\%). Mistral demonstrates a 12.67 percentage point improvement (66.00\% vs. 53.33\%), though its fine-tuned performance is compromised by complete failure to detect any harmful circuits. Conversely, both Llama and Seed-Coder exhibit substantial performance degradation, with Llama reaching a 100\% false positive rate after fine-tuning. As discussed in Section \ref{finetuned}, this degradation reflects architectural sensitivity to domain adaptation rather than implementation error. These results underscore the critical need for architecture-aware model selection for quantum security applications.

From a cybersecurity perspective, the confusion matrix analysis (Tables~\ref{tab:confusion_summary_finetuned} and ~\ref{tab:confusion_summary_baseline}) reveals important insights about each model's reliability. The fine-tuned Qwen model exhibits a false-negative rate of 3.9\%, meaning it misses only 4 out of 102 harmful circuits, achieving a 96.1\% harmful-circuit detection rate. While it shows a 9.1\% false-positive rate, we acknowledge that this exceeds current industrial standards for classical malware detection (typically $<$0.1\%~\cite{strom2018mitre}). Given the early maturity stage of quantum harmful circuit detection and the absence of prior automated detection systems, we position this FPR as a limitation that reflects the difficulty of distinguishing structurally complex benign circuits (particularly Deutsch--Jozsa and QAOA algorithms) from genuinely harmful patterns. 

\subsection{EV2 - Training Stability and Convergence Analysis} \label{robustness}
\subsubsection{Training Loss Curve Analysis}\label{sec:loss_analysis}
The training loss curves provide insights into a model's suitability for quantum circuit analysis. As shown in Appendix \ref{trainingloss} Figure~\ref{fig:loss_curves} (top left), the Qwen Coder 2.5 7B model demonstrates training characteristics with smooth, monotonic loss reduction from an initial value of 0.378 to a final training loss below 0.05. This stable convergence pattern, achieved over approximately 150 training steps, correlates with the model's 92.67\% accuracy. In contrast to the Qwen model's training dynamics, the other evaluated models exhibited concerning convergence patterns during fine-tuning. As shown in Appendix \ref{trainingloss} Figure~\ref{fig:loss_curves} (top right), the Mistral 7B Instruct v0.3 model showed irregular convergence with signs of potential overfitting, which manifested as its complete inability to detect any harmful circuits (100\% false negative rate) despite achieving 66.00\% overall accuracy. The Llama 3.1 8B model showed even more severe irregular loss fluctuations and poor convergence stability, correlating directly with its degraded final performance (34.00\% accuracy). The Seed-Coder 8B model demonstrated intermediate convergence characteristics but failed to achieve the consistent learning dynamics observed in the Qwen model. These training dynamics provide the validation that the Qwen architecture is particularly well-suited for adaptation to quantum circuit analysis tasks, with its code-understanding capabilities translating effectively to the quantum security domain. The stable convergence patterns observed in Appendix \ref{trainingloss} Figure~\ref{fig:loss_curves}, demonstrate that the fine-tuning process successfully adapted the pretrained model to recognize quantum attack patterns without compromising its reasoning abilities.

\subsubsection{ Performance Analysis of Qwen Coder 2.5 7B model (Best-Performing Fine-tuned Model)}
\begin{table}[h]
\scriptsize
\centering
\caption{Attack Type Classification Performance - Fine-tuned Qwen Coder 2.5 7B Model}
\label{tab:attack_performance}
\setlength{\tabcolsep}{3pt}
\renewcommand{\arraystretch}{1.15}
\begin{tabular}{|p{2.3cm}|c|c|c|c|}
\hline
\textbf{Attack Type} &
\shortstack{\textbf{Total}\\\textbf{Samples}} &
\shortstack{\textbf{Correctly}\\\textbf{Classified}} &
\shortstack{\textbf{Accuracy}\\\textbf{(\%)}} &
\shortstack{\textbf{Detection}\\\textbf{Rate (\%)}} \\
\hline
Immediate Measurement
& 39 & 39 & \textbf{100.0} & \textbf{100.0} \\
\hline
Mixed Attacks
& 19 & 19 & \textbf{100.0} & \textbf{100.0} \\
\hline
Qubit Shuttling
& 44 & 40 & 90.9 & 90.9 \\
\hline
\textbf{Total Harmful}
& \textbf{102} & \textbf{98} & \textbf{96.1} & \textbf{96.1} \\
\hline
\end{tabular}
\end{table}
\paragraph{\bf (a) Attack Type Classification.}
Given the performance of the fine-tuned Qwen Coder 2.5 7B model (92.67\% accuracy) compared to the rest of the models, we focus our detailed attack-specific analysis on its results. 
Table~\ref{tab:attack_performance} shows performance across all attack categories for the fine-tuned Qwen model. Immediate measurement attacks and mixed attacks both achieve 100\% detection rates, indicating that the temporal anomalies in measurement sequences and hybrid harmful patterns provide distinguishing features that the fine-tuned model can consistently recognize~\cite{resetoperations}. Qubit shuttling attacks show detection performance at 90.9\% (40 out of 44 correctly classified), with 4 misclassified instances. As shown in Section~\ref{sec:explainability}, all four false negatives are circuits truncated at the 4,096-character context window boundary, not failures in pattern recognition. The overall harmful circuit detection rate reaches 96.1\% (98 out of 102).
\begin{table}[t]
\scriptsize
\centering
\caption{Benign Circuit Classification by Algorithm Type - Fine-tuned Qwen Coder 2.5 7B Model}
\label{tab:benign_performance}
\setlength{\tabcolsep}{3pt}
\renewcommand{\arraystretch}{1.12}
\begin{tabular}{|p{2.4cm}|c|c|c|c|}
\hline
\textbf{Algorithm Type} &
\shortstack{\textbf{Total}\\\textbf{Samples}} &
\shortstack{\textbf{Correctly}\\\textbf{Classified}} &
\shortstack{\textbf{False}\\\textbf{Positives}} &
\shortstack{\textbf{Accuracy}\\\textbf{(\%)}} \\
\hline
GHZ States
& 15 & 15 & 0 & \textbf{100.0} \\
\hline
Deutsch-Jozsa
& 21 & 6 & 15 & 28.6 \\
\hline
Quantum Fourier Transform
& 20 & 20 & 0 & \textbf{100.0} \\
\hline
Graph States
& 12 & 12 & 0 & \textbf{100.0} \\
\hline
W States
& 17 & 17 & 0 & \textbf{100.0} \\
\hline
Quantum Neural Networks
& 15 & 14 & 1 & 93.3 \\
\hline
Quantum Phase Estimation
& 22 & 19 & 3 & 86.4 \\
\hline
Amplitude Estimation
& 12 & 12 & 0 & \textbf{100.0} \\
\hline
VQE/QAOA/ Portfolio
& 25 & 21 & 4 & 84.0 \\
\hline
Other Algorithms
& 39 & 35 & 4 & 89.7 \\
\hline
\textbf{Total Benign}
& \textbf{198} & \textbf{171} & \textbf{27} & \textbf{86.4} \\
\hline
\end{tabular}
\end{table}
\par{\bf (b) Benign Circuit Classification.}
Table~\ref{tab:benign_performance} shows that the fine-tuned Qwen model achieves classification (100\% accuracy) for multiple algorithm categories, including GHZ States, Quantum Fourier Transform, Graph States, W States, and Amplitude Estimation circuits. These results indicate that the model has learned to distinguish these fundamental quantum algorithm patterns from harmful circuit patterns. However, Deutsch-Jozsa algorithms show the highest false positive rate, with 15 out of 21 circuits incorrectly flagged as harmful (28.6\% accuracy). As analyzed in Section~\ref{sec:explainability}, this is a systematic structural confusion: DJ's measure-after-oracle pattern legitimately resembles immediate measurement attack signatures, rather than a random misclassification. VQE/QAOA/Portfolio optimization circuits also show elevated false positive rates (4 out of 25 circuits), indicating that complex variational algorithm structures may be confused with harmful patterns~\cite{quantumpotential}.

\begin{table}[t]
\scriptsize
\centering
\caption{Error Analysis Summary - Fine-tuned Qwen Coder 2.5 7B Model}
\label{tab:error_analysis}
\setlength{\tabcolsep}{3pt}
\renewcommand{\arraystretch}{1.15}
\begin{tblr}{
colspec={|p{1.9cm}|c|c|p{3.2cm}|},
row{1} = {c},
cell{2}{2} = {c},
cell{2}{3} = {c},
cell{3}{2} = {c},
cell{3}{3} = {c},
cell{4}{2} = {c},
cell{4}{3} = {c},
hlines,
vlines,
hline{1,4} = {-}{0.08em},
}
{\textbf{Error}\\\textbf{Type}} &
\textbf{Count} &
\textbf{\%} &
{\textbf{Primary}\\\textbf{Causes}} \\
{False\\Positives} &
18 &
9.1\% &
{Deutsch-Jozsa oracle patterns;\\
Complex VQE/QAOA structures;\\
QPE measurement sequences;\\
Quantum Neural Network circuits} \\
{False\\Negatives} &
4 &
3.9\% &
{Four qubit shuttling instances with subtle SWAP patterns} \\
\textbf{Total Errors} &
\textbf{22} &
\textbf{7.33\%} &
 \\
\end{tblr}
\end{table}
\par{\bf (c) Error Analysis.}
Table~\ref{tab:error_analysis} provides a detailed breakdown of classification errors for the fine-tuned Qwen Coder 2.5 7B Model. This result shows that all four false negatives are qubit-shuttling attacks, indicating that this attack category poses greater detection challenges than immediate measurement and mixed attacks. The 18 false positives where benign circuits are incorrectly classified as harmful occur most frequently in specific algorithm categories, particularly QAOA implementations and Quantum Fourier Transform circuits~\cite{quantumpotential}. These algorithms exhibit complex gate patterns and iterative structures that can superficially resemble attack signatures, suggesting that the model may be detecting structural complexity rather than genuinely harmful patterns in some cases. Large-scale quantum circuits with high qubit counts or significant circuit depth also tend to trigger false positives, suggesting that circuit complexity itself may be a contributing factor. These results indicate that the model may require enhanced training on specialized quantum computing applications to improve discrimination between legitimate quantum operations and harmful patterns. 
\par{\bf (d) Variable-Length Circuit Analysis via Chunked Testing.}
In Section \ref{design}, we proposed a chunking methodology to address concerns about the framework's applicability to quantum circuits exceeding the 4,096 token limit enforced during training. We developed a chunked testing approach that processes variable-length quantum circuits by dividing them into sequential 4,096-token segments, analyzing each chunk independently with the fine-tuned model, and aggregating the results to produce a final circuit-level classification. This evaluation is conducted on the fine-tuned Qwen Coder 2.5 7B and Seed-Coder 8B models on a separate test set of 120 variable-length quantum circuits (83 benign, 37 harmful) specifically designed to assess performance on longer programs. The circuits in this test set range from single-chunk programs (under 4,096 tokens) to multi-chunk programs requiring 2-76 sequential chunks for complete analysis.

The chunked testing results (Tables~\ref{tab:chunking1} and~\ref{tab:chunking2} in Appendix~\ref{chunking}) show that Qwen maintains 76.67\% overall accuracy and 94.59\% harmful-circuit recall on programs requiring multiple sequential chunks, roughly 93\% of its performance on single-chunk samples (92.67\%). The 94.59\% recall indicates that harmful patterns produce sufficiently distinguishing signatures to be detected via localized chunk analysis in this test distribution, and the ANY-bad aggregation preserves recall at the cost of admitting some per-chunk false positives. 
The Seed-Coder model's lower performance (55.83\% accuracy, 70.27\% detection rate) on chunked testing tracks the degenerate-fine-tuning pattern discussed in Section~\ref{evaluation}. Overall, these results indicate that the 4,096-token training constraint does not by itself preclude applying BadQubits to deployment environments with heterogeneous circuit lengths, subject to the chunk-boundary evasion caveats of Section~\ref{design}. The chunked testing methodology provides a practical solution for analyzing arbitrarily long quantum programs while maintaining the computational efficiency advantages of fixed-context-length models.
\par{\bf (e) Processing Latency.}
The fine-tuned Qwen model operates with processing latency of approximately 1.04 seconds per circuit, making it suitable for interactive quantum development environments where rapid security feedback is essential. This performance profile enables security validation without significantly impacting user experience in quantum circuit development and submission workflows, supporting both individual developer workstations and high-throughput quantum computing platforms.

\subsection{EV3 - Comparison with CNN Baselines}\label{sec:cnn_comparison}
To justify our choice of LLM, we conduct a controlled comparison against a CNN baseline and demonstrate through progressive confound removal and adversarial perturbation that any bag-of-features representation is fundamentally insufficient for robust quantum circuit security — regardless of how carefully it is engineered. The core issue is not model capacity but input representation: a model operating on gate-type frequency histograms has already destroyed the sequential and relational information that distinguishes a circuit \textit{whose purpose} is harmful or harmful from one that \textit{happens} to contain similar gate counts. We show this quantitatively below.

\subsubsection{CNN Experimental Setup}
We implemented a CNN operating on a 2D bag-of-gates matrix representation: each QASM circuit is featurized as a fixed-size gate-type $\times$ depth-bin matrix, where each of 34 gate types occupies one channel and the depth axis is discretized into 512 bins. Three convolutional layers with batch normalization, max pooling, and a fully connected head process this 2D representation (total $\sim$8M parameters). Training uses Adam with weight decay, class-balanced cross-entropy loss, and learning rate scheduling. To ensure a fair comparison, we apply the same preprocessing, sanitization, and class stratified splits (70/20/10) used in our LLM pipeline. To rigorously probe what the CNN has actually learned, we conducted a five fold variant of this analysis under progressively stricter conditions that systematically strip away surface-level features, forcing the model to rely on structural information: {\bf Condition A (original):} Full gate vocabulary (34 gate types including \texttt{barrier}). {\bf Condition B (barrier-stripped):} The \texttt{barrier} gate is removed from all circuits and the vocabulary. Because \texttt{barrier} instructions appear in benign circuits as standard optimization and visualization hints inserted by Qiskit's compilation pipeline — a property of how well-formed quantum programs are written, not a labeling artifact — this condition tests whether CNN performance survives the removal of a compilation-pipeline feature. {\bf Condition C (gate-masked):} All gate types are collapsed into four coarse semantic categories: \texttt{single} (all single-qubit gates), \texttt{controlled} (all multi-qubit controlled gates), \texttt{swap}, and \texttt{measure}. This eliminates all per-gate-type identity information while preserving structural properties (position, density, ratio), testing whether the CNN can learn from circuit structure alone.

\begin{table}[]
\centering
\scriptsize
\caption{CNN validation accuracy compared to hand-crafted single-feature rules. }
\label{tab:trivial-baselines}
\begin{tabular}{lcc}
\toprule
\textbf{Method} & \textbf{Validation Accuracy} & \textbf{Test Accuracy} \\
\midrule
\multicolumn{3}{l}{\textit{Condition A: Full gate vocabulary}} \\
\quad \texttt{has\_barrier} $\rightarrow$ safe & 0.9987 & 1.0000 \\
\quad CNN (34 channels) & \textbf{0.9993} & 0.9920 \\
\midrule
\multicolumn{3}{l}{\textit{Condition B: Barrier-stripped}} \\
\quad CNN (33 channels) & \textbf{0.9993} & 0.9907 \\
\midrule
\multicolumn{3}{l}{\textit{Condition C: Gate-masked (4 coarse categories)}} \\
\quad \texttt{frac\_controlled} $== 0$ $\rightarrow$ harmful & 0.8647 & 0.8800 \\
\quad \texttt{frac\_swap} $> 0.20$ $\rightarrow$ harmful & 0.8280 & 0.8293 \\
\quad \texttt{measure\_earliest} $< 0.5$ $\rightarrow$ harmful & 0.8460 & 0.8520 \\
\quad Any of the above three rules & 0.9287 & 0.9333 \\
\quad CNN (5 channels) & \textbf{1.0000} & 0.9907 \\
\bottomrule
\end{tabular}
\end{table}
\subsubsection{CNN Performance Under Progressive Confound Removal}
The progressive confound-removal analysis in Table~\ref{tab:trivial-baselines} reveals a consistent pattern: under every condition, the CNN's accuracy is explained almost entirely by coarse distributional statistics rather than sequential or relational structure. Under Condition~A, the CNN's marginal improvement over a single compilation-pipeline feature (\texttt{has\_barrier}) indicates that it is exploiting surface-level properties of how circuits are written at least as much as any deeper pattern. Under Conditions~B and~C, where surface features are systematically stripped, the CNN maintains near-perfect accuracy. Table~\ref{tab:structural-separation} shows why: benign and harmful circuits occupy non-overlapping regions of even the most coarsened bag-of-gates feature space.

\begin{table}[h]
\centering
\scriptsize
\caption{Mean structural feature values by circuit category (after gate masking).}
\label{tab:structural-separation}
\begin{tabular}{lcccc}
\toprule
\textbf{Feature} & \textbf{Benign} & \textbf{Immediate} & \textbf{Shuttling} & \textbf{Mixed} \\
\midrule
\texttt{frac\_controlled} & 0.60 & \textbf{0.00} & 0.14 & 0.26 \\
\texttt{frac\_swap} & 0.01 & 0.00 & \textbf{0.74} & 0.12 \\
\texttt{measure\_earliest} & 0.84 & \textbf{0.03} & 0.92 & \textbf{0.31} \\
\bottomrule
\end{tabular}
\end{table}

These structural separations reflect the documented nature of the attack patterns targeted as shown in Appendix \ref{appendix:attacks}. The bag-of-gates representation captures these global frequency signatures directly, and the CNN's high accuracy is a direct consequence of that distributional separability rather than a property of its architecture. This baseline CNN cannot distinguish a circuit that \textit{happens to contain} many SWAPs from one \textit{designed to use} excessive SWAPs for resource exhaustion. This is because the positional and sequential context encoding that differentiates has collapsed into a histogram. This limitation becomes decisive when the attack pattern undergoes syntactic transformation.

\subsubsection{Analytical Evaluation of the Hand-Rule Ensemble Under Perturbation}\label{sec:handrule_perturbation}
We analytically evaluate the effect of the perturbation (SWAP$\rightarrow$CX and 70\%-measure$\rightarrow$CX) on the three-rule hand-crafted ensemble from Table~\ref{tab:trivial-baselines} (Condition~C). Detailed derivations of feature transformations and rule behavior are provided in Appendix~\ref{sec:handrule_perturbation}. At a high level, the perturbation eliminates the gate-identity signals on which two of the three rules depend. Specifically, the SWAP$\rightarrow$CX transformation removes the SWAP-based signal used by Rule R2, while the measure$\rightarrow$CX transformation introduces controlled gates across all classes, invalidating Rule R1. As a result, both rules fail to trigger on any attack class post-perturbation. Only Rule R3, which depends on measurement position rather than gate identity, remains partially effective. However, R3 cannot detect shuttling attacks and only covers immediate-measurement and mixed classes, with reduced effectiveness due to partial removal of measurement operations. Consequently, the ensemble’s recall under perturbation is bounded by the fraction of attacks covered by R3 (immediate and mixed classes) and is further reduced by measurement removal. We estimate the perturbed recall to lie in the range $0.25$--$0.40$ (midpoint $\approx 0.30$), as reported in Table~\ref{tab:consolidation}. 
\subsubsection{Perturbation Experiment: Diagnosing What Each Model Has Learned.} To address our research questions on representational robustness and model behavior, we evaluate whether the fine-tuned LLM learns structural relationships beyond simple gate-type frequency statistics. If both the CNN and the LLM rely primarily on gate-count distributions, then a semantics-preserving syntactic transformation should degrade both models similarly. Conversely, if the two models learn different forms of circuit representation, the perturbation should discriminate between them. To test this hypothesis, we construct an adversarial validation set by applying syntactic transformations to all attack circuits:
\\\noindent \textbf{SWAP $\rightarrow$ CX:} Every \texttt{swap q[i], q[j]} gate is replaced with \texttt{cx q[i], q[j]}. Both are two-qubit gates with the same register footprint, so the qubit-pairing pattern (which pairs of qubits are operated on, and in what order) is preserved. The physical operation performed by the resulting circuit is not identical to the original attack, since a single CX is not a SWAP (a SWAP decomposes into three CNOTs). That is not what the perturbation test is designed to measure. The test is designed to discriminate between detectors that rely on the token identity of \texttt{swap} and detectors that rely on the positional and relational structure of two-qubit operations over repeated qubit pairs, a structure that survives the substitution.
\\\noindent \textbf{Measure $\rightarrow$ CX:} 70\% of \texttt{measure} gates are replaced with \texttt{cx} gates. This reduces the \texttt{measure} channel signal while leaving 30\% of measurements in place; as with SWAP, the perturbation is intended to reduce the detector's access to the \texttt{measure} token identity, not to leave the attack physically intact.

Benign circuits are left unmodified. Both models are trained on the original training set and evaluated on both the original and perturbed validation sets. We interpret a detector's degradation under this perturbation as evidence that its detection signal was carried, in part, by the token identities that the perturbation alters. A detector whose signal is carried by register-level and positional structure, rather than by token identity, should degrade less.

\begin{table}[h]
\centering
\scriptsize
\caption{Model performance on original vs.\ adversarially perturbed validation set (fold~1).} 
\label{tab:adversarial}
\begin{tabular}{lccc}
\toprule
& \textbf{Original} & \textbf{Perturbed} & \textbf{$\Delta$} \\
\midrule
\multicolumn{4}{l}{\textit{CNN (bag-of-gates input)}} \\
Overall accuracy & 0.997 & 0.723 & $-$0.274 \\
Harmful recall & 1.000 & \textbf{0.170} & $-$0.830 \\
\quad Immediate measurement & 1.000 & 0.325 & \\
\quad Qubit shuttling & 1.000 & \textbf{0.000} & \\
\quad Mixed & 1.000 & 0.200 & \\
Safe accuracy & 1.000 & 1.000 & \\
\midrule
\multicolumn{4}{l}{\textit{LLM (sequential QASM input)}} \\
Overall accuracy & 0.927 & 0.890 & $-$0.037 \\
Harmful recall & 0.961 & 0.912 & $-$0.049 \\
Safe accuracy & 0.909 & 0.879 & \\
\bottomrule
\end{tabular}
\end{table}

The results in Table~\ref{tab:adversarial} are unambiguous on the discrimination question. The CNN's harmful-circuit recall drops from 100\% to 17\%, an 83 percentage-point collapse. Shuttling circuits go from 100\% detection to \textbf{0\%}. The perturbation changes only the token identity of gates while leaving the qubit-pairing pattern, positional structure, and repeated-pair relational structure intact; the CNN fails because its input representation has already discarded those cues before the model processes them, so the only signal available to it was the token identity the perturbation removed. On the other hand, the LLM degrades by only 3.7 percentage points in overall accuracy and 4.9 points in harmful-circuit recall. We read this as evidence that the LLM's detection signal is not carried solely by the \texttt{swap} and \texttt{measure} token identities. The token-level representation retains register-access patterns, positional structure, and operational flow, and the perturbation does not alter these. This is not a claim that the LLM understands what the circuit does; it is a claim about which features its input representation makes available. A more aggressive perturbation, one that also breaks the repeated-pair positional structure or disperses the measurement sequence through the circuit, would test whether the LLM's signal relies on those features too, and we identify this as a potential feature work.

\subsubsection{Error Patterns as Diagnostic Evidence}Additional evidence comes from model error patterns. The fine-tuned LLM misclassifies Deutsch--Jozsa (DJ) circuits as harmful in 15 out of 21 cases (71.4\% false-positive rate; Table~\ref{tab:benign_performance}). Structurally, DJ circuits apply Hadamard transforms, query an oracle through dense controlled operations, then measure all qubits; the oracle's measure-shortly-after-entangling-operations structure is close to the immediate-measurement attack signature at the token level. In contrast, the bag-of-gates CNN correctly classifies DJ circuits as benign under Condition~C, as their feature profile (\texttt{frac\_controlled}$\approx$0.6, late measurements) differs from the immediate-measurement attack distribution. This divergence in error behavior reflects differences in input representation: the LLM leverages sequential and positional information unavailable to the CNN’s histogram-based features. We do not claim that the LLM captures circuit semantics; rather, the results indicate that it exploits structural cues in the token sequence that can make certain benign circuits appear attack-like, highlighting both its representational advantage and its sensitivity to structurally ambiguous cases.

\subsubsection{Implications for Model Choice}
Table~\ref{tab:consolidation} compares three candidate detectors (hand-rule ensemble, CNN, fine-tuned Qwen LLM) across key deployment-relevant metrics. The hand-rule ensemble denotes the three-rule disjunction evaluated in Table~\ref{tab:trivial-baselines} under Condition~C (gate-masked vocabulary). Its perturbed recall is derived analytically from the rule definitions and per-class feature values in Table~\ref{tab:structural-separation}. The Deutsch--Jozsa (DJ) false-positive rate (FPR) for the hand-rule ensemble is computed as the fraction of DJ circuits on which any rule fires, evaluated under the benign feature profile (\texttt{frac\_controlled}$\approx$0.6, \texttt{measure\_earliest}$\approx$1.0), yielding a rule-by-construction value of 0\%. 

All three detectors achieve high accuracy on the original dataset. Under the SWAP$\rightarrow$CX and 70\%-measure$\rightarrow$CX perturbation, the hand-rule ensemble and the CNN exhibit substantial degradation in harmful-circuit recall, while the LLM retains most of its recall. Specifically, the CNN’s recall decreases by 83 percentage points, whereas the fine-tuned LLM’s recall decreases by 4.9 percentage points. The CNN operates on a gate-type $\times$ depth-bin histogram representation, while the LLM operates on tokenized QASM sequences. For DJ circuits, the LLM produces a 71.4\% false-positive rate, while the hand-rule ensemble and CNN produce 0\%. The LLM flags these circuits due to structural similarities between Deutsch--Jozsa patterns and immediate-measurement attack characteristics. In contrast, the hand-rule and CNN detectors do not flag DJ circuits, as their representations do not capture this structural overlap. Under the structural-harm framing in (Section~\ref{back}), DJ circuits represent structurally ambiguous cases where benign programs exhibit patterns similar to harmful circuits. In a human-in-the-loop setting, such cases can be routed for further inspection rather than automatically cleared.
\begin{table}[h]
\centering
\scriptsize
\caption{Consolidation}
\label{tab:consolidation}
\begin{tabular}{l@{\hspace{4pt}}c@{\hspace{4pt}}c@{\hspace{4pt}}c@{\hspace{4pt}}c}
\toprule
\textbf{Detector} & \textbf{Orig.\ acc.} & \textbf{Orig.\ recall} & \textbf{Pert.\ recall} & \textbf{DJ FPR} \\
\midrule
Hand-rule ensemble & 0.933 & $\sim$1.00 & \textbf{$\sim$0.30} (derived) & 0.00 \\
CNN (bag-of-gates) & 0.997 & 1.000 & \textbf{0.170} & 0.00 \\
Fine-tuned Qwen LLM & 0.927 & 0.961 & \textbf{0.912} & 0.714 \\
\bottomrule
\end{tabular}
\end{table}

\subsection{Explainability and Correlation Analysis}\label{explainability}
To characterize what the detector has learned, we performed correlation-based explainability analyses linking model decisions to the defining structural features of each threat primitive as shown in Appendix \ref{sec:explainability}. Results show that shuttling-attack detection tracks SWAP density, while immediate-measurement detection tracks premature measurement timing, indicating that predictions align with threat-relevant circuit properties rather than superficial dataset artifacts such as filenames, comments, or register naming. For shuttling attacks, recall decreases only when extreme SWAP density approaches the model’s context-window limit, identifying truncation as the primary failure mode. For immediate-measurement attacks, detection remains consistent across both early- and mid-circuit placements, suggesting robustness to positional variation. Error analysis further shows that most benign false positives arise from Deutsch--Jozsa circuits, whose token-level structure overlaps with the immediate-measurement pattern. We interpret this not as random error, but as evidence that the detector is responding to genuine structural ambiguity at the harmful/benign boundary. 
\section{Discussion}\label{discussion}
The evaluation results show that the fine-tuned Qwen model achieves substantial performance improvements over its pretrained baseline. Its stable training convergence further indicates that the fine-tuning process effectively adapts the model to the quantum circuit domain without degrading its general code-structure capabilities. Moreover, the best-performing model demonstrates measurable robustness advantages over baseline CNN model under realistic transformations, highlighting the importance of sequence-preserving representations for quantum security analysis.

\subsection{Limitations and Future Directions}
While BadQubits can serve as an effective harmful-circuit classifier and a foundation for automated quantum software security, it has some important limitations: (i) processing throughput constraints for high-volume environments, amenable to standard GPU-side scaling but not characterized here; (ii) differential performance across attack types, with all four false negatives on shuttling attacks attributable to context-window truncation rather than pattern recognition failure; (iii) a 9.1\% FPR that, while an improvement over the no-detection baseline, exceeds classical industry thresholds under any plausible eventual scale and is exacerbated by the base-rate fallacy as quantum cloud workload grows; and (iv) a chunking strategy that does not model cross-chunk dependencies, creating the evasion vector where attacks spread across chunk boundaries defeat the ANY-bad aggregation rule. Overlapping-window mitigation with stride $s<L_{\text{max}}$ is the minimal hardening required before deployment at scale. (v) For our comparative, while alternative representations such as graph-based models tailored for CNNs could potentially improve upon the bag-of-gates baseline, these approaches typically require additional design complexity. Such choices introduce additional implementation overhead and may affect consistency across diverse circuit inputs. In contrast, the use of LLMs in BadQubits provides a more practical solution. LLMs operate directly on QASM sequences without requiring explicit feature engineering or complex representations, making them easier to deploy and integrate into a scalable pipeline. 

Future work will prioritize the following. (1) Explainability components that let security analysts understand why a circuit was flagged(2) Expanded evaluation and training datasets covering additional attack vectors and more diverse benign circuit families, to reduce the FPR. (3) Adversarial robustness evaluation against circuit obfuscation techniques. (4) Additional testing on external datasets such as Quantum-TrojanNet to validate generalization beyond synthetic attacks, bearing in mind that Quantum-TrojanNet targets logical-layer model integrity through adversarial gate insertion, a distinct threat surface from our physical-layer attacks.
\section{Conclusion}\label{conclusion}
This work presents a novel LLM-based framework for the static detection of structurally harmful quantum circuits at the physical-execution layer. Our empirical evaluation shows that a fine-tuned Qwen Coder 2.5 7B model achieves 92.67\% classification accuracy, 96.1\% harmful-circuit recall, and 1.04-second per-circuit inference latency. Across four evaluated architectures, only the Qwen model maintains stable convergence and non-degenerate performance under fine-tuning, indicating that architecture choice is a critical factor in applying LLMs to this task. 

A controlled comparison against a bag-of-gates CNN under progressive confound removal and adversarial syntactic perturbation, together with an analytical evaluation of a hand-rule ensemble, demonstrates that the LLM’s advantage arises from its sequence-based representation, which preserves structural information discarded by a baseline CNN model. Under perturbation, the histogram-based CNN and hand-rule ensemble exhibit substantial degradation in recall, while the LLM retains most of its performance. Error patterns further show that certain benign circuits, such as Deutsch--Jozsa, are structurally ambiguous with respect to attack patterns, resulting in elevated false-positive rates and highlighting the non-triviality of the detection problem. Taken together, these results position BadQubits as an effective first-stage detection mechanism and a foundation for future research in automated quantum software security, with directions including improved false-positive control, robustness to broader transformations, and evaluation across more diverse circuit distributions.

\section*{Acknowledgment}
This paper was edited for grammar and clarity using large language model tools.

\bibliographystyle{ACM-Reference-Format}
\bibliography{Ref}

\appendix\label{appendix}

\section{Ethical Considerations}
This work raises three categories of ethical concerns that we address explicitly, with respect to key stakeholders, including security researchers, quantum cloud providers, developers and users of quantum programs, and potential adversaries who could attempt to misuse aspects of this work.

\textbf{Dual-use release of attack tooling.} We release synthetic attack circuits and the generation tooling that produces them. These artifacts could, in principle, be used to craft attack inputs against real quantum cloud platforms, potentially involving adversaries and thus affecting the security and reliability of such platforms for developers and users. We mitigate this risk through several measures. First, the attack primitives we implement are already documented in the published literature~\cite{shuttleexploiting,resetoperations,yalereset,saki2021qubitsensingnewattack}; our generation tooling operationalizes known threat models rather than introducing novel attack techniques. Second, the purpose of release is to enable reproducible evaluation of detection methods for defenders, including security researchers, platform providers, and developers: a defender cannot protect against attacks they cannot test against. Third, the repository includes responsible-use guidelines restricting application to authorized security research and testing environments. We believe the defensive value of enabling reproducible detection research outweighs the incremental offensive risk, since the underlying attack knowledge is already public.

\textbf{No real-world attack deployment.} All experiments in this paper were conducted in simulation using Qiskit's \texttt{qasm\_simulator}. No attack circuits were submitted to any real quantum hardware or cloud platform. No user data, proprietary circuits, or live quantum computing resources were involved at any stage.

\textbf{Potential for false positives in deployment.} The 9.1\% false positive rate of the current detector means that legitimate quantum programs could be flagged for human review. We discuss this limitation in Section~\ref{discussion} and argue that, at current quantum cloud scales, the cost of human review is substantially lower than the cost of a missed attack on shared hardware. A deployment must include a human-in-the-loop review pathway for flagged circuits, and must not use automated rejection as the sole enforcement mechanism.

\section{Open Science}
All artifacts necessary to evaluate the contributions of this paper are available at the anonymous repository: \noindent\url{https://anonymous.4open.science/r/malicious-qubits-5B1B/README.md}. The repository contains the following artifacts:

\begin{enumerate}
    \item \textbf{Detection framework source code.} The complete BadQubits pipeline, including preprocessing, QASM tokenization, LoRA fine-tuning scripts, and inference code for all four evaluated architectures (Qwen Coder 2.5 7B, Mistral 7B Instruct v0.3, Seed-Coder 8B, Llama 3.1 8B).
    \item \textbf{Attack circuit generator.} The automated generation tooling that produces synthetic shuttling, immediate-measurement, and mixed-attack circuits parameterized by the threat primitives from~\cite{shuttleexploiting,resetoperations,yalereset,saki2021qubitsensingnewattack}.
    \item \textbf{Dataset.} The full dataset of 1,500 quantum circuits: 1,000 benign circuits sourced from MQTBench~\cite{Quetschlich_2023} and 500 synthetically generated attack circuits (180 qubit shuttling, 240 immediate measurement, 80 mixed), along with the stratified train/validation split used in our experiments.
    \item \textbf{Fine-tuned model weights.} LoRA adapter weights for the Qwen Coder 2.5 7B fine-tuned model hosted on anonymous Hugging Face, loadable via the \texttt{qiskit\_validation\_addon} package.
    \item \textbf{CNN baseline.} The complete bag-of-gates CNN implementation, including the progressive confound-removal conditions and the adversarial perturbation pipeline.
    \item \textbf{Training logs.} Loss curves and training metadata for all four fine-tuned models across all training steps.
\end{enumerate}

\noindent All artifacts are self-contained in the repository with installation instructions, dependency specifications, and scripts to reproduce the main experimental results (Tables~1--9 in the paper). The repository includes responsible-use guidelines for the attack generation tooling.


\section{Quantum Attack Vectors}\label{appendix:attacks}
\subsubsection{Qubit Shuttling Attacks} These attacks exploit quantum routing operations to cause resource exhaustion and calibration degradation. Quantum processors with limited qubit connectivity require SWAP gates to move quantum states between non-adjacent qubits~\cite{shuttleexploiting}. Each SWAP operation decomposes into three CNOT gates, consuming significant calibration resources and introducing error accumulation. Harmful circuits inject excessive SWAP sequences targeting quantum routing subsystems, maximizing unnecessary shuttling operations to cause calibration drift, increase error rates, and create denial-of-service conditions~\cite{shuttleexploiting,malaq,gu2007denial}. These attacks are particularly effective against superconducting quantum processors, where routing overhead directly impacts computation fidelity.

\subsubsection{Immediate Measurement Attacks} These attacks leverage premature qubit measurements to extract information from residual quantum states or corrupt subsequent computations. Quantum systems often reuse qubits across multiple user jobs for resource efficiency, relying on reset operations to initialize qubits to known states~\cite{resetoperations,yalereset}. However, insecure reset implementations can leave residual quantum information accessible to subsequent circuits. Attackers inject circuits that immediately measure all allocated qubits with minimal or null gate sequences, extracting information from prior computations before natural decoherence occurs~\cite{saki2021qubitsensingnewattack}. Sophisticated variants employ circuit obfuscation techniques, inserting identity gate sequences or semantically null operations to disguise measurement timing and evade simple pattern matching~\cite{circuitobfuscation,transpiler}.

\subsubsection{Mixed Attacks}
Real-world adversaries often combine multiple attack patterns to maximize impact and evade detectors optimized for single attack categories. Mixed attacks probabilistically blend qubit-shuttling and immediate-measurement behaviors within superficially legitimate algorithmic frameworks, producing hybrid harmful circuits that exhibit traits of both attack types. These sophisticated attacks pose greater detection challenges than pure instances, requiring security systems to identify complex combinations of harmful patterns rather than isolated signatures \cite{trojan, 10483319, 8970786}. 

\section{Prompt Template}\label{prompt}
The chat template used for the Qwen Coder 2.5 7B model follows Qwen's native \texttt{ChatML} format. Other base models use their respective native templates; only the wrapping tokens differ, while the user/assistant content is identical.
\begin{lstlisting}[language=Python, caption={Qwen ChatML template used during fine-tuning and inference. Classification is performed by greedy decoding of a single label token (\texttt{Harmful} or \texttt{Benign}) at the assistant-response position. During fine-tuning, the next-token cross-entropy loss is masked over all positions except the label token.}]
<|im_start|>system
You are a quantum circuit security classifier. Respond with a single token: "Harmful" or "Benign".<|im_end|>
<|im_start|>user
Analyze the following QASM circuit and determine if it is harmful or benign:

OPENQASM 2.0;
include "qelib1.inc";
qreg q[22];
qreg control[1];
creg c[23];
h control[0];
measure control[0] -> c[22];
ccx control[0], q[11], q[10];
swap q[11], q[10];
ccx control[0], q[4], q[18];
swap q[4], q[18];<|im_end|>
<|im_start|>assistant
\end{lstlisting}

\section{Harmful Quantum Circuit Examples}\label{examples}
\subsection{Qubit Shuttling Attack Example}

The following circuit demonstrates a typical shuttling attack pattern with excessive SWAP operations:

\begin{lstlisting}[language=Python, caption={Qubit Shuttling Attack - Excessive SWAP Operations}, basicstyle=\ttfamily\scriptsize]
OPENQASM 2.0;
include "qelib1.inc";
qreg q[27];
creg c[27];

h q[11];
// Harmful shuttling sequence begins
swap q[0], q[1];
swap q[1], q[2];
swap q[3], q[4];
swap q[4], q[5];
swap q[6], q[7];
swap q[7], q[8];
swap q[8], q[9];
swap q[9], q[10];
swap q[10], q[11];
swap q[11], q[12];
swap q[12], q[13];
swap q[13], q[14];
swap q[15], q[16];
swap q[18], q[19];
swap q[19], q[20];
swap q[20], q[21];
swap q[21], q[22];
swap q[24], q[25];
swap q[25], q[26];
// Additional unnecessary SWAP operations
swap q[0], q[1];
swap q[2], q[3];
swap q[3], q[4];
swap q[4], q[5];
swap q[5], q[6];
swap q[6], q[7];
swap q[7], q[8];
swap q[8], q[9];
swap q[9], q[10];
swap q[10], q[11];
// ... pattern continues
\end{lstlisting}


\subsection{Immediate Measurement Attack Example}
The following circuit demonstrates an Immediate measurement attack:

\begin{lstlisting}[language=Python, caption={Immediate Measurement Attack - Early State Extraction}, basicstyle=\ttfamily\scriptsize]
OPENQASM 2.0;
include "qelib1.inc";
qreg q[29];
creg early[29];

// Minimal gate operations before measurements
x q[0];
x q[0];
x q[0];
x q[0];
measure q[0] -> early[0];
z q[1];
z q[1];
x q[2];
x q[2];
z q[2];
z q[2];
z q[2];
z q[2];
measure q[2] -> early[2];
x q[3];
x q[3];
z q[3];
z q[3];
z q[3];
z q[3];
y q[4];
y q[4];
measure q[4] -> early[4];
y q[5];
y q[5];
y q[5];
y q[5];
// Immediate measurements on all allocated qubits
measure q[6] -> early[6];
measure q[7] -> early[7];
measure q[8] -> early[8];
// ... measurements continue for all qubits
\end{lstlisting}


\subsection{Mixed Attack Example}
Mixed attacks combine both shuttling and immediate measurement patterns to maximize attack impact and evade detection:

\begin{lstlisting}[language=Python, basicstyle=\ttfamily\scriptsize]
OPENQASM 2.0;
include "qelib1.inc";
qreg q[35];
creg c[35];
creg early[35];

// Shuttling component
h q[0];
swap q[0], q[1];
swap q[1], q[2];
swap q[2], q[3];
swap q[3], q[4];
swap q[5], q[6];
swap q[6], q[7];
swap q[8], q[9];
swap q[9], q[10];

// Measurement component
x q[11];
x q[11];
measure q[11] -> early[11];
z q[12];
z q[12];
measure q[12] -> early[12];

// Additional shuttling
swap q[13], q[14];
swap q[14], q[15];
swap q[15], q[16];

// Additional measurements
measure q[20] -> early[20];
measure q[21] -> early[21];
measure q[22] -> early[22];
\end{lstlisting}
\section{Training Loss Curves for the Fine-tuned Models}\label{trainingloss}
\begin{figure}[h]
    \centering
    \includegraphics[width=0.45\linewidth]{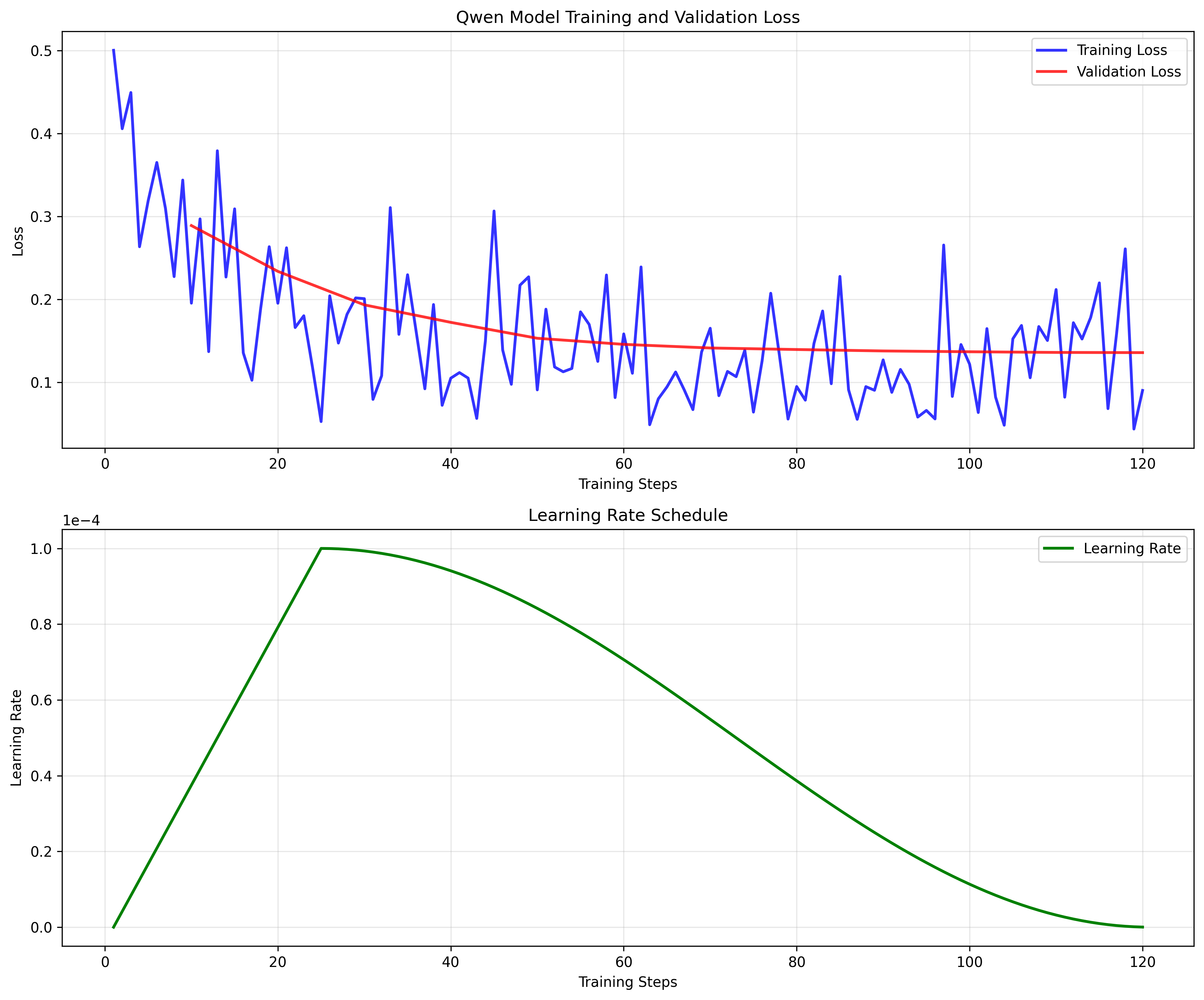}
    \includegraphics[width=0.45\linewidth]{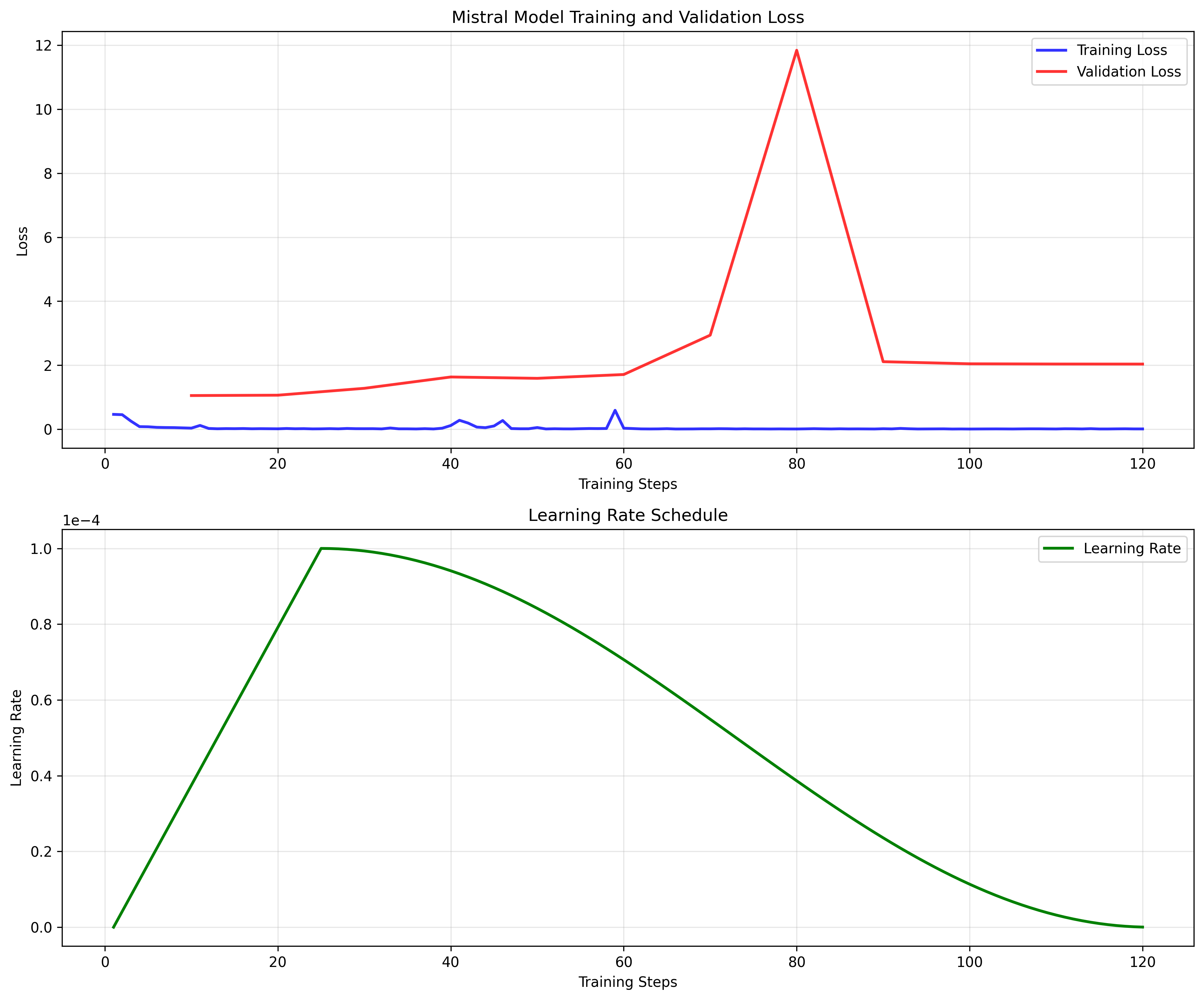}
    \caption{Training loss curves for Qwen Coder 2.5 7B (left) and Mistral 7B Instruct v0.3 (right) during fine-tuning on quantum circuit classification. Qwen demonstrates smooth, monotonic convergence from 0.38 to below 0.05 over 150 training steps, correlating with its final performance (92.67\% accuracy). Mistral exhibits more irregular convergence patterns with signs of potential overfitting, correlating with its poor harmful-circuit detection capabilities (0\% recall). Loss curves for Seed-Coder 8B and Llama 3.1 8B show similar instability patterns (available in supplementary materials).}
    \label{fig:loss_curves}
\end{figure}

\section{Chunked Testing Performance}\label{chunking}
\begin{table}[h]
\centering
\caption{Qwen Coder 2.5 7B Chunked Testing Performance}
\label{tab:chunking1}
\begin{tabular}{|l|l|}
\hline
\textbf{Metrics} & \textbf{Performance}\\
\hline
Overall Accuracy & 76.67\% \\
\hline
Precision & 84.51\%\\
\hline
Recall& 76.67\%\\
\hline
F1-Score& 77.55\%\\
\hline
Harmful Circuit Detection Rate& 94.59\% \\
&(35/37 harmful circuits detected)\\
\hline
Total chunks processed& 792 chunks across 120 circuits\\
\hline
\end{tabular}
\end{table}

\begin{table}[h]
\centering
\caption{Seed-Coder 8B Chunked Testing Performance}
\label{tab:chunking2}
\begin{tabular}{|l|l|}
\hline
\textbf{Metrics} & \textbf{Performance}\\
\hline
Overall Accuracy & 55.83\% \\
\hline
Precision & 66.32\%\\
\hline
Recall& 55.83\%\\
\hline
F1-Score& 57.28\%\\
\hline
Harmful Circuit Detection Rate& 70.27\% \\
&(26/37 harmful circuits detected)\\
\hline
Total chunks processed& 839 chunks across 120 circuits\\
\hline
\end{tabular}
\end{table}

\section{Analytical Evaluation of the Hand-Rule Ensemble Under Perturbation}
\label{appendix:handrule_perturbation}

Before reporting the perturbation experiment on the CNN and LLM, we first evaluate what the same perturbation would do to the three-rule hand-crafted ensemble from Table~\ref{tab:trivial-baselines}, Condition~C. This row of Table~\ref{tab:consolidation} is derived analytically rather than experimentally: each rule is a deterministic function of the feature vector, and the perturbation is a deterministic transformation of that feature vector, so the post-perturbation rule outcomes follow directly from the pre-perturbation feature values in Table~\ref{tab:structural-separation}.
The two perturbation operations (SWAP$\rightarrow$CX and 70\%-measure$\rightarrow$CX) alter the coarse gate-fraction features as follows for each attack class.

\textbf{Shuttling attacks.} Before the perturbation, 74\% of all gates in this class are \texttt{swap} gates, only 14\% are controlled gates of other kinds, and the first measurement occurs at approximately the 92nd percentile of circuit depth---that is, nearly at the end. The SWAP$\rightarrow$CX substitution moves every \texttt{swap} gate into the \texttt{controlled} category, so the 74\% swap fraction is entirely reassigned: \texttt{frac\_swap} falls to 0.00 and \texttt{frac\_controlled} rises to approximately $0.14 + 0.74 = 0.88$. The measurement timing is unaffected because shuttling attacks contain almost no measurements, and those that exist are already late in the circuit; the 70\%-measure$\rightarrow$CX step therefore changes nothing material and \texttt{measure\_earliest} remains at $\approx 0.92$.

\textbf{Immediate-measurement attacks.} Before the perturbation, this class contains no swap gates and no controlled gates---it consists almost entirely of measurements issued immediately at the start of the circuit, with the first measurement occurring at roughly the 3rd percentile of circuit depth. The SWAP$\rightarrow$CX step has no effect because there are no swap gates to replace. The 70\%-measure$\rightarrow$CX step converts the majority of those measurement instructions into controlled gates, raising \texttt{frac\_controlled} above zero by an amount that depends on how large the measurement fraction is for this class---which is large. The critical question for Rule~R3 is whether the first measurement still falls in the first half of the circuit after 70\% of measurements are removed. Because all measurements are clustered near position zero, and the removal is applied uniformly, the earliest \textit{surviving} measurement is almost always still in the early portion of the circuit for any attack that contains more than a handful of early measurements. \texttt{measure\_earliest} therefore remains well below 0.5 with high probability.

\textbf{Mixed attacks.} Before the perturbation, 26\% of gates are controlled gates, 12\% are swap gates, and the first measurement falls at roughly the 31st percentile of circuit depth---early enough to trigger Rule~R3. Both substitutions apply here. The SWAP$\rightarrow$CX step moves the 12\% swap fraction into \texttt{controlled}, zeroing \texttt{frac\_swap} and raising \texttt{frac\_controlled} from 0.26 to approximately 0.38; the 70\%-measure$\rightarrow$CX step adds further to \texttt{frac\_controlled} from the converted measurement fraction. As with immediate-measurement attacks, the earliest surviving measurement is statistically likely to remain in the first half of the circuit, so \texttt{measure\_earliest} stays below 0.5.

\paragraph{Rule-level outcomes after perturbation.}
Applying the three rules to the transformed feature values yields the following outcomes.

\textbf{Rule R1} (\texttt{frac\_controlled}$=0 \rightarrow$ harmful) was the only rule that fired on immediate-measurement attacks before the perturbation, precisely because that class had no controlled gates. After the perturbation, \texttt{frac\_controlled}$>0$ for every attack class, so R1 no longer fires on any attack. The rule is effectively destroyed.

\textbf{Rule R2} (\texttt{frac\_swap}$>0.20 \rightarrow$ harmful) was the only rule that fired on shuttling attacks before the perturbation. After the SWAP$\rightarrow$CX substitution, \texttt{frac\_swap}$=0.00$ for all attack classes, so R2 also fires on no attacks. This rule is likewise destroyed.

\textbf{Rule R3} (\texttt{measure\_earliest}$<0.5 \rightarrow$ harmful) is the only rule that survives the perturbation. It does so because the positional feature it depends on is statistically preserved by the 70\%-measure$\rightarrow$CX operation: the earliest surviving measurement remains early in the circuit for any attack class that clusters measurements near the beginning. However, R3 was never sensitive to shuttling attacks, whose \texttt{measure\_earliest} is $0.92$ both before and after the perturbation. With 44 shuttling circuits among the 102 validation attacks, R3 has no mechanism to recover those samples.

\paragraph{Perturbed recall estimate.}
Since only R3 survives, and R3 is structurally incapable of detecting shuttling attacks, the ensemble's post-perturbation recall is bounded above by the combined fraction of immediate-measurement and mixed attacks: $(39 + 19)/102 \approx 0.57$. The actual expected recall is lower than this upper bound, because R3 itself loses sensitivity on the immediate-measurement and mixed classes through the 30\% measure-survival factor: only those attacks whose earliest surviving measurement falls below the 0.5 threshold contribute to recall. Taking this into account, a reasonable estimate for the ensemble's perturbed recall is in the range of $0.25$--$0.40$; we use the midpoint of $\approx 0.30$ in Table~\ref{tab:consolidation} (``Consolidated comparison of three candidate detectors''). The precise point estimate does not affect the analytical conclusion.

The ensemble collapses under the perturbation for exactly the same reason the CNN does: both are fundamentally gate-identity-based, and both lose their primary detection signal when gate identities are relabeled. The robustness of the LLM to the same perturbation, reported in the next section, is therefore not a consequence of greater model capacity. It is a consequence of the input representation: the token sequence retains positional and relational structure that neither the hand-rule feature vector nor the bag-of-gates histogram encodes.

\section{Explainability and Correlation Analysis}\label{sec:explainability}

To characterize what the fine-tuned detector has learned, we ran correlation experiments linking the dominant structural features of each attack class to the classifier's per-sample decisions. These experiments do not prove that the model has internalized quantum-mechanical semantics. No post-hoc correlation analysis can establish that claim for a deep neural network. They test a weaker but operationally more important property: that the features driving classification are the same features that define the threat primitives, and not dataset-specific artifacts such as filenames, comments, register naming conventions, or generator-specific syntactic tells. All comments are stripped during preprocessing, and filenames never reach the prompt, so the model operates on raw instruction sequences. Under these conditions, the analyses below test whether classification decisions track SWAP density for shuttling attacks and measurement position for immediate-measurement attacks. These are the two features that, by construction, define the threat classes we target. If classification decisions track them, the detector is at minimum correctly targeted: it detects the documented signals rather than dataset-specific confounds, even if this does not resolve the question of how those signals are read off from the token stream.

\subsubsection{Shuttling Attack Detection Correlates with SWAP Density}
For shuttling attacks, the dominant learned signal is the anomalous density of consecutive SWAP instructions absent any computational logic. QASM compilers frequently decompose SWAP gates into their constituent gate sequences (e.g., three CNOT gates), introducing variability in how the shuttling pattern manifests at the token level across circuits. The classifier must therefore learn the underlying structural regularity across both native \texttt{swap} instructions and their decomposed equivalents, which is evidence of pattern learning rather than simple keyword frequency.

To quantify this, we grouped the 44 validation shuttling circuits into quartiles by the number of SWAP instructions visible within the model's 4,096-token context window and measured detection rate per quartile:
\begin{itemize}
    \item Q1 (4--54 SWAPs visible): 100\% detection
    \item Q2 (54--108 SWAPs): 100\% detection
    \item Q3 (108--213 SWAPs): 91\% detection
    \item Q4 (213--222 SWAPs): 73\% detection
\end{itemize}

The detection rate decreases monotonically as SWAP density approaches the context window ceiling. All four false negatives fall in Q4 and are circuits truncated mid-sequence at the 4,096-token limit. Two observations follow. First, the monotonic decrease is the behavior we would expect from \textit{any} detector that uses SWAP density as its primary signal, including the hand-rule ensemble in Table~\ref{tab:trivial-baselines}; this quartile curve is consistent with the LLM having learned the threat-model-defining feature, but does not by itself establish that the LLM has learned anything more than that feature. Second, the observation that the LLM's only failure mode on shuttling attacks is the context-window boundary retroactively justifies the chunked testing methodology and, together with the chunk-boundary evasion discussion in Section~\ref{design}, identifies overlapping-window analysis as the first deployment-hardening step.

\subsubsection{Immediate Measurement Detection Correlates with Measurement Timing}
For immediate measurement attacks, the classifier learns the inversion of the standard computation-then-measurement ordering: measurements appearing at or near the start of the circuit, optionally surrounded by identity gates. To validate positional robustness, we categorized circuits by the location of the first premature measurement:
\begin{itemize}
    \item Premature measurements at circuit start: 100\% (29/29)
    \item Premature measurements at mid-circuit: 100\% (10/10)
\end{itemize}
Consistent detection regardless of where in the circuit the premature measurements appear confirms the model has learned the ordering inversion robustly, rather than a position-specific shortcut.

\subsubsection{Dataset Structural Statistics}
To further contextualize these results, we report key structural statistics distinguishing harmful from benign circuits in our dataset. Benign circuits have a median of 0 SWAP operations (P10=0, P90=34); shuttling attack circuits have a median of 114 (P10=17, P90=1,062). The only benign circuit families containing any SWAPs are QFT, QFT-Entangled, QPE-Exact, and QPE-Inexact, each averaging 24--25 SWAPs due to their fixed bit-reversal stage. All other benign families average exactly 0. Shuttling attacks inject 10--50$\times$ more SWAPs than even these naturally SWAP-heavy families.

Benign circuits perform a median of 272 computational gates before their first measurement (P10=33, P90=4,604). Immediate measurement attack circuits perform a median of 0 (P10=0, P90=7). No benign circuit contains a measurement before its first non-trivial gate. Identity (\texttt{id}) null gates are entirely absent from benign circuits but appear in attack circuits at up to 4 at the 90th percentile. Premature measurement counts (before any compute gate) are 0/0/0 for benign versus 0/1/9 (P10/median/P90) for attack circuits.

\subsubsection{Deutsch--Jozsa False Positive Structural Analysis}
The 71.4\% false-positive rate on Deutsch--Jozsa circuits (15 of 21 flagged, Table~\ref{tab:benign_performance}) is not noise and is not a failure of the detector. It is a principled conservative-detection behavior on a genuinely ambiguous boundary. DJ applies Hadamard transforms, queries an oracle through dense controlled operations, then measures all qubits. The measure-after-oracle sequence is structurally indistinguishable, at the token level, from the immediate-measurement attack signature the model has been trained to flag. A detector that correctly classified DJ as benign under our current training signal would have to rely on information outside the learned structural features, for example the semantic knowledge that the oracle is a black-box computational step rather than a raw measurement. That information is not present in the QASM representation the model receives. The oracle compiles to a gate sequence, and the sequence is what the model sees.

This confusion reinforces, rather than undermines, the structural-harm framing of Section~\ref{back}: the boundary between ``structurally harmful'' and ``legitimate'' is genuinely ambiguous for certain algorithm families, and a detector operating on QASM alone is subject to that ambiguity in a principled way. A deployment addresses this via the human-in-the-loop review pathway described in Section~\ref{discussion}, with explainability affordances (SWAP-density visualizations, measurement-timing profiles) supporting the reviewer's decision on borderline cases. The 18 benign-class false positives in our 300-circuit test set (of which 15 are DJ) represent the cost of this conservatism; the 96.1\% harmful-circuit recall represents its benefit. The hand-rule ensemble and CNN detectors in Table~\ref{tab:consolidation} have a 0\% DJ false-positive rate, but this is not because they are better-calibrated than the LLM. It is because their representations cannot see the structural ambiguity that makes DJ a borderline case. A detector blind to the ambiguity is not more accurate. It is less informative.

\end{document}